\documentclass[structabstract,natbib]{aa}  

\newcommand{\sse}{{\tt sourceExtractorSussextractor}}
\newcommand{\bnd}{{\tt sourceExtractorTimeline}}
\newcommand{\hers}{{\it Herschel}}
\newcommand{\mic}{$\mu$m}
\usepackage{natbib}
\usepackage{graphicx}
\usepackage{txfonts}
\usepackage{array}
\usepackage{color}
\usepackage{rotating}
\usepackage[colorlinks=true,linkcolor=blue,citecolor=blue,urlcolor=blue]{hyperref}

\bibpunct{(}{)}{;}{a}{}{,} 

\begin{document}

\title{The {\it Herschel}\thanks{{\it Herschel} is an ESA space observatory with science instruments provided by a European-led principal investigator consortia and with an important participation from
NASA.} Virgo Cluster Survey}

\subtitle{XIX. Physical properties of low luminosity FIR sources at $z <$ 0.5}

\author{Ciro Pappalardo\inst{1,2}, Luca Bizzocchi\inst{3}, Jacopo Fritz\inst{4}, Alessandro Boselli\inst{5}, Mederic Boquien\inst{10, 14}, Samuel Boissier\inst{5}, Maarten Baes\inst{6}, Laure Ciesla\inst{7}, Simone Bianchi\inst{8}, Marcel Clemens\inst{9}, Sebastien Viaene\inst{6}, George J. Bendo\inst{11}, Ilse De Looze\inst{6, 10, 12}, Matthew W. L. Smith\inst{13}, Jonathan Davies\inst{13}}
\institute{
  Centro de Astronomia e Astrof\'{\i}sica da Universidade de Lisboa,  Observat\'{o}rio Astron\'{o}mico de Lisboa, Tapada da Ajuda, 1349-018 Lisboa, Portugal
    \and
  Instituto de Astrof\'{\i}sica e Ciencias do Espa\c{c}o, Universidade
de Lisboa, OAL, Tapada da Ajuda, PT1349-018 Lisboa, Portugal
   \and
   Max-Planck-Institute f\"{u}r Extraterrestrial Physik, Giessenbachstrasse 1, D-85748 Garching bei M\"{u}nchen
    \and
 Instituto de Radioastronom\'{i}a y Astrof\'{i}sica, CRyA, UNAM, Campus Morelia, A.P. 3-72, C.P. 58089, Michoac\'{a}n, Mexico
  \and
  Aix Marseille Universit\'e, CNRS, LAM (Laboratoire d'Astrophysique de Marseille) UMR 7326, 13388, Marseille, France
       \and
   Sterrenkundig Observatorium, Universiteit Gent, Krijgslaan 281 S9, B-9000 Gent, Belgium
    \and
 Department of Physics, University of Crete, 71003, Heraklion, Greece
  \and
 Osservatorio Astrofisico di Arcetri – INAF, Largo E. Fermi 5, 50125 Firenze, Italy
  \and
  Osservatorio Astronomico di Padova, Vicolo dell'Osservatorio 5, I-35122 Padova, Italy. 
\and  
  Institute of Astronomy, University of Cambridge, Madingley Road, Cambridge, CB3 0HA, UK
   \and
   UK ALMA Regional Centre Node, Jodrell Bank Centre for Astrophysics, School
of Physics and Astronomy, The University of Manchester, Oxford Road,
Manchester M13 9PL, UK
\and
 Dept. of Physics \& Astronomy, University College London, Gower Street, London WC1E 6BT, UK 
 \and
 Department of Physics and Astronomy, Cardiff University, The Parade, Cardiff, CF24 3AA, UK
 \and
Unidad de Astronomia, Facultad de Ciencias Basicas, Universidad de Antofagasta, Avenida Angamos 601, Antofagasta, Chile
}

\date{}
\abstract
    {The star formation rate is a crucial parameter for the investigation galaxy evolution. At low redshift the cosmic star formation rate density declines smoothly, and massive active galaxies become passive, reducing their star formation activity. This implies that the bulk of the star formation rate density at low redshift is mainly driven by low mass objects.} 
    {We investigate the properties of a sample of low luminosity far-infrared sources selected at 250 $\mu$m. We have collected data from ultraviolet to far-infrared in order to perform a multiwavelengths analysis. The main goal is to investigate the correlation between star formation rate, stellar mass, and dust mass for a galaxy population with a wide range in dust content and stellar mass, including the low mass regime that most probably dominates the star formation rate density at low redshift.
    }
    {We define a main sample of $\sim$ 800 sources with full spectral energy distribution coverage between 0.15 $< \lambda <$ 500 $\mu$m and an extended sample with $\sim$ 5000 sources in which we remove the constraints on the ultraviolet and near-infrared bands. We analyze both samples with two different spectral energy distribution fitting methods: MAGPHYS and CIGALE, which interpret a galaxy spectral energy distribution as a combination of different simple stellar population libraries and dust emission templates.}
    {In the star formation rate versus stellar mass plane our samples occupy a region included between local spirals and higher redshift star forming galaxies. These galaxies represent the population that at $z <$ 0.5 quenches their star formation activity and reduces their contribution to the cosmic star formation rate density. The subsample of galaxies with the higher masses ($M_\ast >$ 3 $\times$ 10$^{10}$ M$_\odot$) do not lie on the main sequence, but show a small offset as a consequence of the decreased star formation. Low mass galaxies ($M_\ast <$ 1 $\times$ 10$^{10}$ M$_\odot$) settle in the main sequence with star formation rate and stellar mass consistent with local spirals.} 
  {Deep \hers\ data allow the identification of a mixed galaxy population with galaxies still in an assembly phase or galaxies at the beginning of their passive evolution. We find that the dust luminosity is the parameter that allow us to discriminate between these two galaxy populations. The median spectral energy distribution shows that even at low star formation rate our galaxy sample has a higher mid-infrared emission than previously predicted.} 
\titlerunning{SED analysis of HeViCS background sources}
\authorrunning{Pappalardo et al.}
\maketitle

\section{Introduction}
\label{intro}

Different studies in the past have addressed the problem of the evolution of the cosmic star formation rate (SFR) across the Hubble time \cite[see][and references therein]{mad2}. The SFR is a crucial parameter for investigating galaxy evolution: from the Big Bang up to $z \sim$ 1-3 during the ``galaxy assembly epoch'', most of galaxies grew and evolved to finally settle into the main sequence, the tight relation observed between the SFR and the galaxy stellar mass ($M_\ast$) \citep{bri,noe,dad,oli2,kar,elb,rod}.
Below $z <$ 1 cosmic SFR density (SFRD) declines smoothly, and massive active galaxies become passive early type with low star formation activity \citep{sand,lil,mad,bos3,red,car}. If massive galaxies shut down their star formation activity, the bulk of the SFRD at low redshift should be mainly driven by low mass objects \citep{gav2,cow,bos3}. \cite{hea} confirmed this idea, showing that galaxies with high masses form their stars earlier and evolve more rapidly than low mass objects \cite[see Fig. 2 in][]{hea}. As a consequence, mass discriminates different evolutionary paths: high mass galaxies ($M_\ast \gtrsim$ 1 $\times$ 10$^{11}$ M$_\odot$) drive the rise of SFRD during the assembly epoch, while low mass galaxies ($M_\ast <$ 3 $\times$ 10$^{10}$ M$_\odot$) have constant SFRs at least since $z \sim$ 3, a result found also in numerical simulations \citep{fon,str,alm}. In other words, if we want to investigate the SFRD at $z <$ 1 we should focus our attention on galaxies with $M_\ast$ below  $\approx$ 1 $\times$ 10$^{11}$ M$_\odot$ because at low redshift this is the actual galaxy population that drives the total SFRD. 

The best way to investigate the reasons for such a trend would then be to estimate stellar mass, gas fraction, and dust emission in a large sample of low mass galaxies, to trace in detail the SFRD decline, and to search for possible correlations with the galaxies growth rate and the gas available to feed the star formation. Thanks to the improved sensitivity of ground based instruments, the stellar component of such a population can be investigated in great detail, detecting galaxies with $M_\ast \sim$ 10$^7$ M$_\odot$ up to $z$ = 2.5 \citep[e.g. CANDELS survey,][]{wuy}. For the other components the task is more complex: observations of gas in low luminosity galaxies even at low redshift is highly time consuming, while for the dust emission there are problems related to the instrument sensitivity. The bulk of dust emission is at far infrared (FIR) wavelengths, a spectral range where observations are severely limited by confusion noise. The advent of the {\it Herschel} Space Observatory \citep{pil} with its two detectors, PACS \citep{pog} and SPIRE \citep{gri}, has opened a new era in the comprehension of galaxy dust properties, allowing a better characterization of the role of dust in star formation processes \citep[see the review of][and references therein]{lut}. 

The goal of this paper is to investigate the evolution of the SFRD at $z <$ 0.5 focusing on the low FIR luminosity galaxies, the main contributors to the SFRD at low redshift.
To achieve this goal we consider one of the largest and deepest {\it Herschel} surveys, the \hers\ Virgo Cluster Survey \citep[][HeViCS]{dav2,dav}. Its main advantage is the sensitivity and the uniformity of data: the 84 square degrees of the survey have been observed with eight orthogonal cross scans, reaching at 250 $\mu$m a depth in flux density close to the confusion noise limit \citep[see][]{pap}. HeViCS has several advantages over other large \hers\ surveys. HerMES \citep{oli} observed 380 square degrees of sky decomposed in different fields of different sizes and with different numbers of cross scans. The data set is not homogeneous, and if we want to use the full coverage of the survey we cannot rely on a FIR selection because different fields will have a different number of sources, above all at low fluxes. A common method of overcoming this problem is to use a {\it Spitzer} 24 $\mu$m counterpart to match the {\it Herschel} sources. However, 24 $\mu$m emission is more sensitive to hot dust and biases the sample towards galaxies with higher SFRs. The other large survey is H-ATLAS \citep{eal}, which observed uniformly 510 square degrees of sky, but the field has been covered with only two cross scans.

With HeViCS we have been able to tackle the galaxy population with low dust content that is still feeding the cosmic SFRD. The {\it Herschel} IR/sub-mm data set has been enriched with complementary data at other wavelengths in order to perform a multiwavelength analysis of its spectral energy distribution (SED). Reproducing the observed emission in such a wide wavelength range by means of theoretical models allows us to derive the physical quantities that are most critical for our analysis (dust mass, star formation rate, stellar mass). The multiwavelength nature of this study, and the fact that the galaxies under analysis have a very good coverage in terms of data points at the various frequencies, makes our sample an ideal one with which to construct an average SED that can be described as representative of low luminosity FIR objects. Such a SED can be used as a benchmark to investigate galaxies at higher $z$ because lower dust temperatures at higher $z$ are found more frequently than previously thought. Different studies indicate a colder dust temperature in high $z$ sub-mm galaxies with respect to local galaxies with similar FIR luminosities \citep{hwa,mag}.

The paper is organized as follows. Sections \ref{hd} and \ref{met} describe the sample and the method used to extract relevant physical parameters. The results are shown in Section \ref{res} together with the properties of the average SEDs. Conclusions are given in Section \ref{conc}.

\begin{figure}\begin{center}
\includegraphics[clip=,width= .49\textwidth]{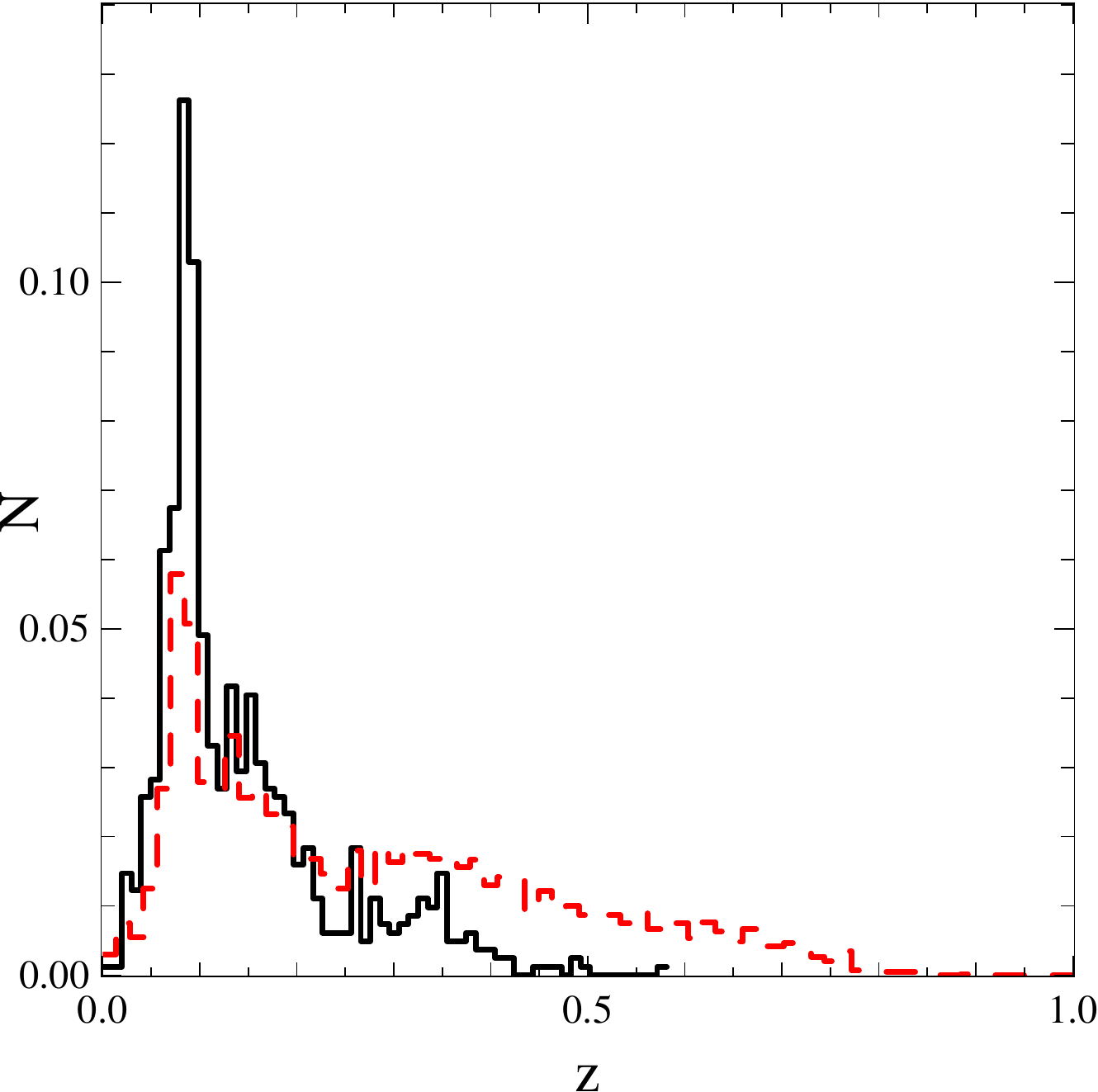}\end{center}
\caption{Normalized redshift distribution of the main (black solid line) and the extended (red dashed line) sample. Sources in both samples are selected to have a counterpart in SDSS, UKIDSS, and WISE with a reliability above 80\%. For the extended sample we removed the constraints on NIR and UV data.}
\label{red}\end{figure}

\section{Sample}
\label{hd}

In this section we explain how we build the sample used for the analysis.

\subsection{Herschel data}

The catalog that we used as a starting point to construct our sample is the point source catalog of \cite{pap} selected at 250 \mic\ from HeViCS. This survey observed about 84 square degrees of sky centered on the Virgo cluster, using both SPIRE \citep{gri} and PACS \citep{pog}, in the spectral domain between 100-500 $\mu$m. Details about SPIRE and PACS data reduction are given in \cite{aul} and \cite{dav}. The main difference in the data reduction, with respect to the standard {\it Herschel} pipelines, is for the SPIRE instrument, for which we used an alternative technique for correcting temperature drifts, the {\it BRIght Galaxy ADaptive Element} (BriGAdE, M. Smith PhD Thesis{\footnote{{\it http://orca.cf.ac.uk/42751/}}}).

PACS data were reduced up to level 1 using HIPE version 10.0.0 \cite{ott}. At this point only the signal drift, the $1/f$ noise, and the detection of glitches still need to be corrected. These tasks are performed by the IDL algorithm {\tt Scanamorphos} \citep{rou}, which also takes care of the map making. The full width at half maximum (FWHM) of the two instruments was 9\arcsec, 13\arcsec, 17 \farcs5, 23\farcs9, and 35\farcs1 at 100, 160, 250, 350, and 500\,\mic, respectively{\footnote{{\it http://herschel.esac.esa.int/twiki/pub/Public/SpireCalibrationWeb\\ /beam\_release\_note\_v1\-1.pdf}}} and the maps had a pixel size of 1\farcs7, 2\farcs85, 6\arcsec, 8\arcsec, and 12\arcsec\ at 100, 160, 250, 350, and 500\,\mic.

\begin{figure}\begin{center}
\includegraphics[clip=,width= .49\textwidth]{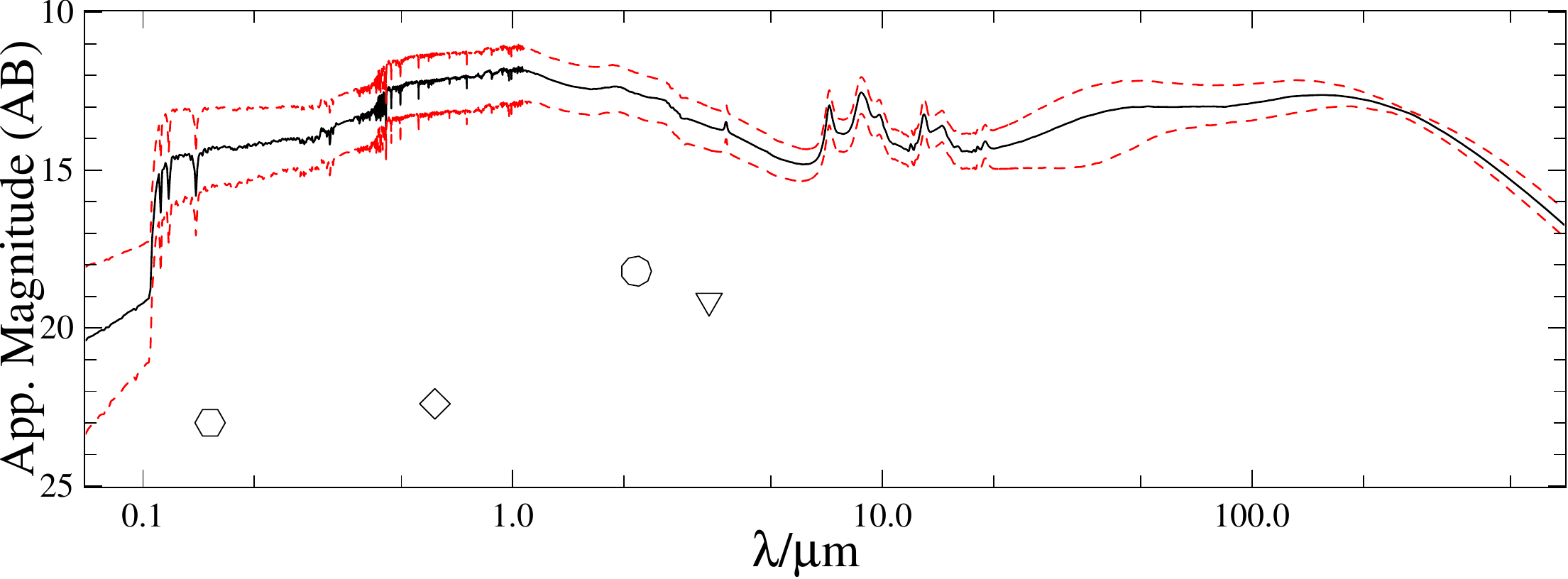}\end{center}
\caption{Median SED of the galaxies in the main sample (black solid line) with the 16th and 84th percentile (red dashed lines). Hexagon, diamond, circle, and triangle show the completeness limits in FUV, $r$, $J$, and W1 band, as given in Sect. \ref{anc}.}
\label{compl}\end{figure}

\begin{figure} \begin{center} 
\includegraphics[clip=,width=.49\textwidth]{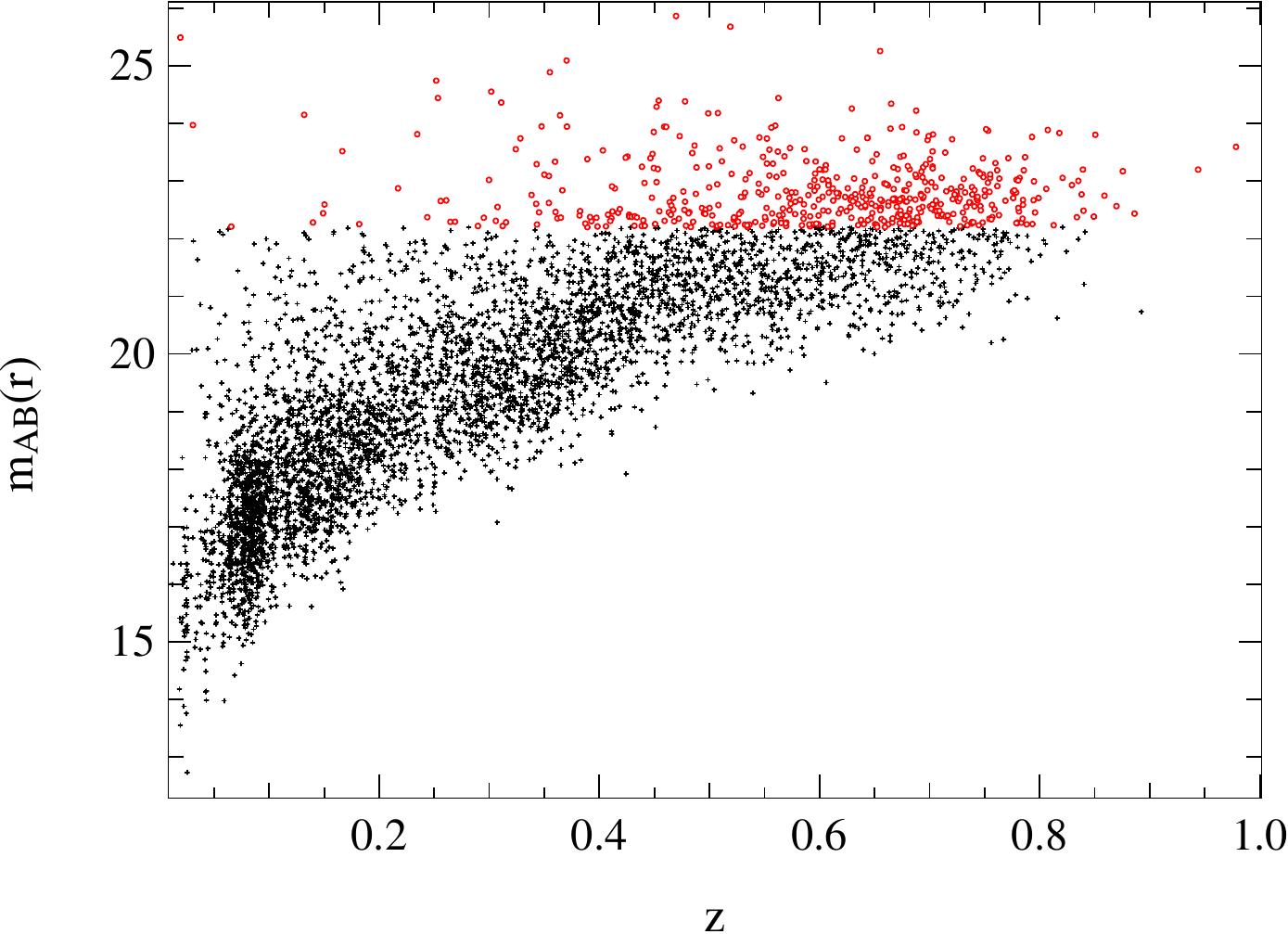}
\includegraphics[clip=,width=.39\textwidth]{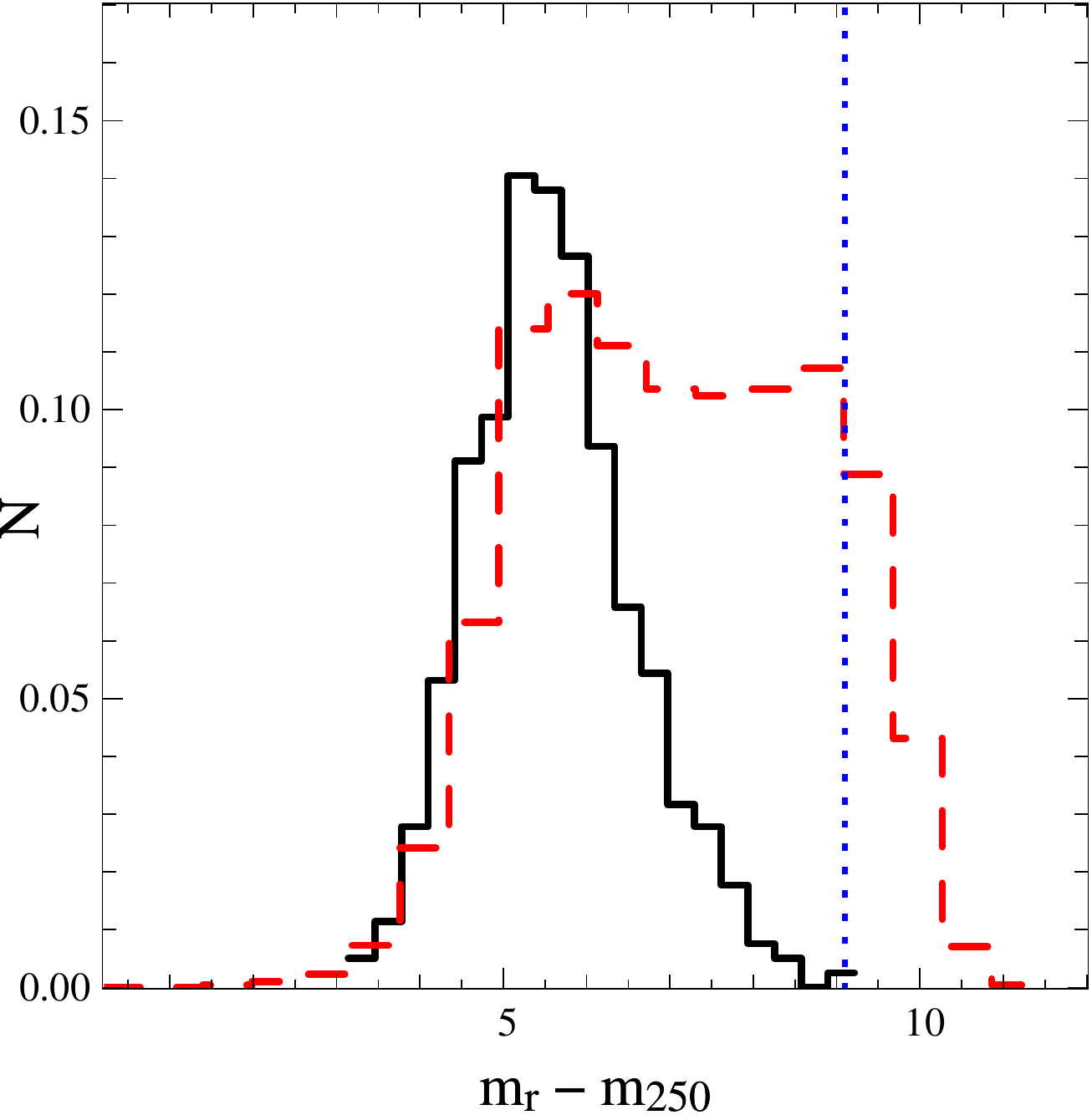}
\end{center} \caption{Top panel: $r$-band AB magnitude as a function of $z$ for the extended sample. Red circles represents the galaxies with $m_r >$ 22.2 excluded from the sample because of incompleteness. Bottom panel: Normalized histograms of the $m_r-m_{250}$ index for the main (black solid line) and extended (red dashed line) sample. The vertical blue dotted line shows the color limit considering both $r$-band and 250 $\mu$m detection limits.}
\label{mrvsz} \end{figure}

The procedure followed to extract the source flux densities from the SPIRE instrument is explained in \cite{pap}. The position of the sources was found using the \sse\ \citep{sav} task in HIPE 10.0.0, considering only sources with flux densities above 20\,mJy. This threshold corresponds to a bit more than 3$\sigma$ above the confusion noise, estimated at 250 \mic\ in 5.8$\pm$0.3\,mJy beam$^{-1}$ \citep{ngu}.

Each SPIRE-detected source was fitted with a Gaussian function using \bnd, a timeline-based point source fitter implemented in HIPE 10.0.0 \citep{ben}. This method fits timeline data from all bolometers within an individual array with a two-dimensional Gaussian function. PACS photometry was estimated using the {\tt pacsAnnularSkyAperturePhotometry} task in HIPE 10.0.0. This procedure performs an aperture photometry on the target at a given radius. The sky is estimated inside a concentric annular radius around the source using an algorithm adapted from {\tt daophot} \citep{ste}. We chose circular apertures with 8\farcs7 and 13\farcs1 of radii at 100 and 160 \mic\ \citep{rig}, and the annular radius between 25\arcsec and 35\arcsec at both wavelengths. According to the HIPE instruction manual \footnote{\it http://herschel.esac.esa.int/hcss-doc-13.0/load/dag/html/Dag.ImageAnalysis.HowTo.AperturePhotometry.html} the error given in {\tt pacsAnnularSkyAperturePhotometry} task is not correct for flux-calibrated \hers\ images. To estimate the photometric errors we calculate the fluxes in eight different apertures at the same distance of the annulus chosen for the sky estimation, and with the same aperture fixed for the sources. The apertures were evenly spaced around the source, and the standard deviation of the fluxes found at these positions gave the error on the measured photometry \citep[see also][]{bal}. 

The methods described above gave a photometric accuracy for PACS and SPIRE data of 15\%, 15\%, 6\%, 11\%, and 21\% at 100, 160, 250, 350, and 500 $\mu$m, respectively \citep[see][for SPIRE bands]{pap}.

\subsection{Ancillary data}
\label{anc}

In addition to \hers, we collected a set of ancillary data at other wavelengths described in the following:

\begin{itemize}
\item far-ultraviolet (FUV) and near-ultraviolet (NUV): The {\it GALEX} Ultraviolet Virgo Cluster Survey \citep[GUViCS,][]{bos} is a survey that explored a region of $\sim$ 120 deg$^2$ in NUV ($\lambda_{eff}$ = 2316 $\AA$, FWHM = 5\farcs6) and 40 deg$^2$ in FUV ($\lambda_{eff}$ = 1539 $\AA$, FWHM = 4\arcsec) centered on M87, the big early-type galaxy in the center of the Virgo cluster. The GUViCS catalog \citep{voy} contains about 1.2 million point-like sources with a 1$\sigma$ error of $\sim$ 0.2 mag down to AB magnitudes of $m_{FUV} \sim$ 23. Photometry was estimated using the {\it MAG\_AUTO} Kron magnitudes \citep{kro} developed in {\it SExtractor} \citep{ber}.
\item Optical data: Photometry in $u$, $g$, $r$, $i$, and $z$ bands come from the SDSS DR10 survey \citep{ahn}. This survey observed about two million sources in the same area as HeViCS  with a completeness limit of 95\% at $m_r \approx 22.2$ AB mag and an average FWHM $\sim$ 1\farcs3. 
\item Near-infrared (NIR):  The UK Infrared Deep Sky Survey Large Area Survey \citep[UKIDSS-LAS,][]{law} has uniformly scanned all the sky in $Y$ ($\lambda_{eff}$ = 1.03 \mic), $J$ ($\lambda_{eff}$ = 1.24 \mic), $H$ ($\lambda_{eff}$ = 1.63 \mic), and $K_s$ ($\lambda_{eff}$ = 2.2 \mic) with an average FWHM $\sim$ 1\farcs2. In this work we used Data Release 10, which observed $\sim$ 2 million objects inside the HeViCS area with a magnitude limit (99\% complete) of 18.2 in $K_s$ band and a typical error of 0.02 mag. In order to improve the quality of the fits we removed arbitrarily the $Y$ and $H$ band and did not consider them in the rest of the analysis. The variation in the estimated physical parameters when considering or not these two bands is, in general, very small; the most relevant is the stellar mass determination. This quantity is best constrained in the NIR, and we found variations of less than 13\%. On the other hand, this choice would allow us to increase the number of reliable fits by $\sim$ 30\%. 
\item Mid-infrared (MIR): The Wide Field Infrared Survey \citep[WISE,][]{wri} is a survey that observed all the sky at four different wavelengths: 3.4 \mic, 4.6 \mic, 12 \mic, and 22 \mic, reaching a 5$\sigma$ sensitivity of 0.08, 0.11,1, and 6 mJy at 3.4, 4.6, 12, and 22 \mic, respectively. We found about 1.7 million of WISE sources in the Virgo Cluster field observed by HeViCS with \hers. Photometry was estimated via PSF-fitting (keyword {\tt wmpro} in the reference catalog \footnote{\it http://irsa.ipac.caltech.edu/cgi-bin/Gator/nph-scan?mission=irsa\&submit=Select\&projshort=WISE}).
\end{itemize}

\begin{figure} \begin{center} 
\includegraphics[clip=,width=.49\textwidth]{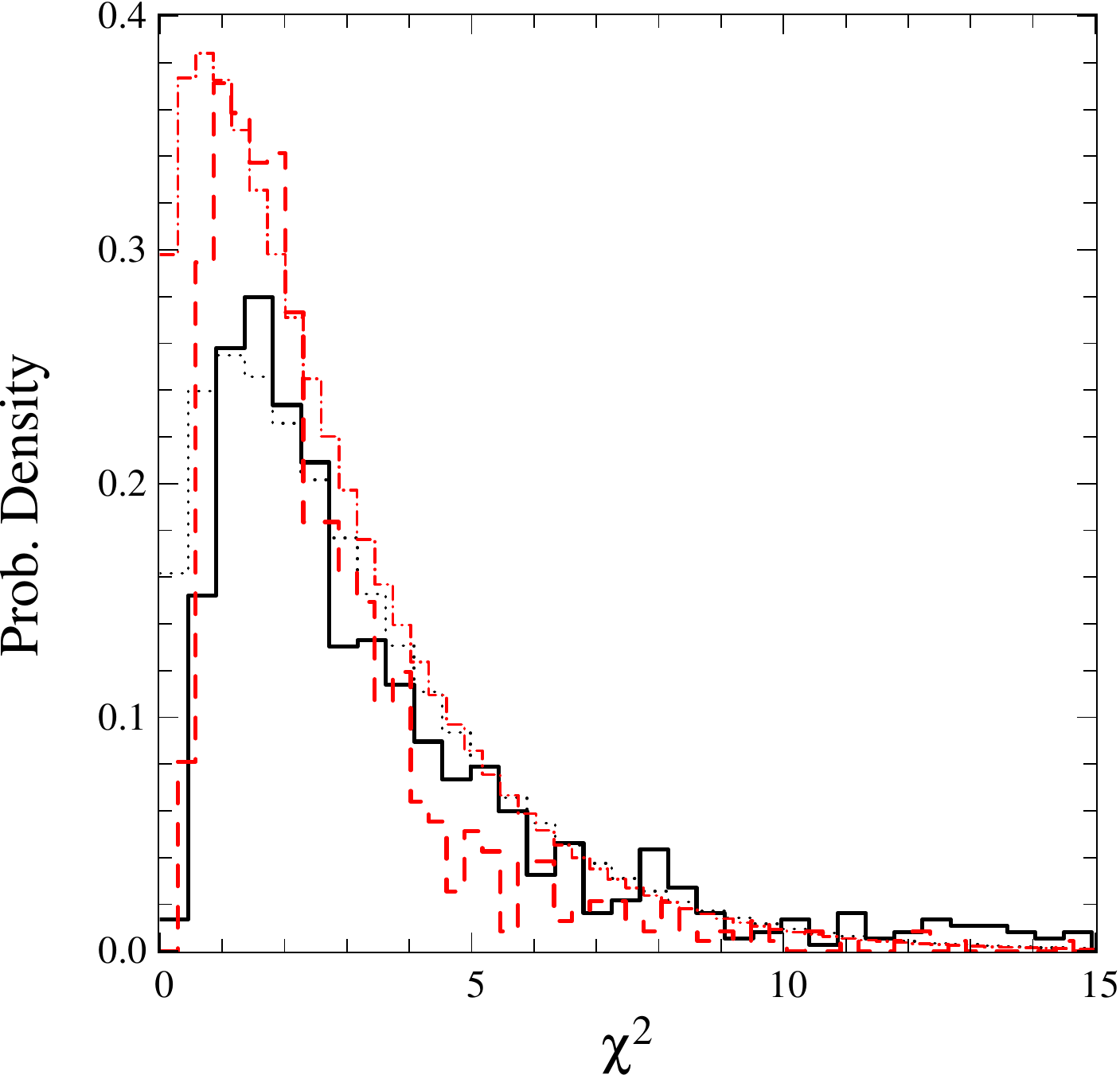}
\end{center} \caption{Probability density function of the main sample obtained using MAGPHYS (black solid line) and CIGALE (red dashed line). Black dotted and Red dot-dashed line show the best fit obtained from Eq. \ref{chif} for MAGPHYS and CIGALE, respectively.}
\label{chi2} \end{figure}

\subsection{Final sample}

We considered all the sources selected at 250 \mic\ in the catalog of \cite{pap}, with flux densities $F_{250} >$ 20 mJy, and S/R$_{250} >$ 3. Since the number of sources in the HeViCS area for SDSS, UKIDSS, and WISE data is so high that the probability of finding a spurious source close to a FIR object is non-negligible, the simple nearest object criterion for finding a counterpart to a FIR source is no longer valid \citep{ric}. We used a likelihood technique developed in  \cite{sut}, used successfully to identify the optical counterparts of \hers\ selected sources \citep{smi2, pap}, and also to cross-correlate H-ATLAS sources with VIKING NIR data \citep{fle}, WISE \citep{bon}, and IRAC \citep{kim}. This method associates a parameter -- the reliability ${\mathcal R}$ -- to each candidate, quantifying the probability that the source considered is the true counterpart of the FIR reference object. The reliability accounts for the possibility that the counterpart is below the  magnitude limit of the reference band or the probability of serendipitously finding a spurious source at some position. As reference magnitudes we chose the $r$, $J$, and $W1$ bands for SDSS, UKIDSS, and WISE respectively. For each HeViCS source the identification with ${\mathcal R_{sdss}}$, ${\mathcal R_{UKIDSS}}$, and ${\mathcal R_{wise}} >$ 80\% was considered sufficient to remove the spurious objects. We did not apply a further cut in SNR in these bands because sources with high reliability have SNRs abundantly above 3. 

The GuViCS sources were cross-correlated with the SDSS selecting the nearest object inside a radius of 5\arcsec\ in \cite{voy}. With this approach 75\% of the sources had multiple SDSS or UV counterparts, but \cite{voy} considered only objects with a 1-1 match in their catalog. At UV wavelengths we then simply matched the objects with the same SDSS identification.

Since we are interested in low luminosity galaxies with moderate star formation we removed stars and/or quasars according to the {\tt SDSS\_phot\_Obj\_Type} keyword in the SDSS catalog and $\sim$ 20 sources classified as active galactic nuclei based on WISE data, a criterion defined in \cite{bon}. We also removed all the sources with $m_r >$ 22.2 to stay 95\% complete in $r$-band. We ended up with 790 sources (hereafter the main sample) with a spectral coverage in 18 different photometric bands. We considered the photometric redshift given in SDSS and, where available, the redshift obtained with spectra ($\sim$ 50\% of the sample). The median redshift is 0.1$\pm _{0.07}^{0.22}$ (16th-84th percentile), shown in Fig. \ref{red}. 

Since we are cross-correlating different data sets, we must check the possibility that a source might be detected at 250 $\mu$m and not at other wavelengths because of the low flux at that band. This case would bias the whole analysis and for this reason needs to be investigated. Fig. \ref{compl} shows the median SED of the galaxies in the main sample normalized to the flux limit imposed at 250 $\mu$m, i.e., 20 mJy (13.1 AB magnitudes), compared to the flux limit of FUV, $r$, $J$, and $W1$ bands, respectively. The magnitude limit in each band is well below the typical SED of our galaxy populations, implying no variation to the completeness level of our catalog due to the cross-correlation at other wavelengths.

Ultraviolet-selected objects undergo a selection effect due to their inclination. Statistically, there are fewer face-on galaxies than inclined ones. For the latter, dust correction is higher, and in extreme cases this can result in a complete absorption of the UV emission. These galaxies will be undetected at UV, even thought star formation is taking place. To take into account this possibility we defined an additional ``extended sample'', in which we remove the constraints on the UV and NIR selections. This selection ends up with $\sim$ 5000 sources with median $z$ = 0.22$\pm _{0.08}^{0.48}$ (16th-84th percentile), shown in Fig. \ref{red}. The higher median redshift requires caution when considering selection effects, more relevant at higher $z$ \citep[see e.g. Fig. 2 in][]{dun}. To quantify this effect in the top panel of Fig. \ref{mrvsz} we compare the $r$-band AB magnitude as a function of $z$. The objects with $m_r >$ 22.2 are less than 10\% of the extended sample. Up to $z$ = 0.6 the faint sources represents only 5\% of the total sample, rising to 43\% for $z >$ 0.6.

Another possibility to consider is that we could still lose highly dust obscured galaxies owing to our optical limit. To investigate this we consider an index color given by the difference in AB magnitudes $m_r-m_{250}$. In the bottom panel of Fig. \ref{mrvsz} we show the histograms of $m_r-m_{250}$ for the main and extended sample. The vertical dotted line shows the color index obtained considering the sensitivity limits in $r$-band and 250 $\mu$m. For the main sample all the galaxies considered are well below this threshold, while for the extended sample 9\% of the objects are outside the color limit. However, most of these sources are at $z >$ 0.6 where we already showed that selection effects become relevant. In other words, our selection allows us to obtain a representative sample of the dusty galaxies for all of the main sample and at least up to $z <$ 0.6 for the extended sample.

\begin{figure}\begin{center}
\includegraphics[clip=,width=.4\textwidth]{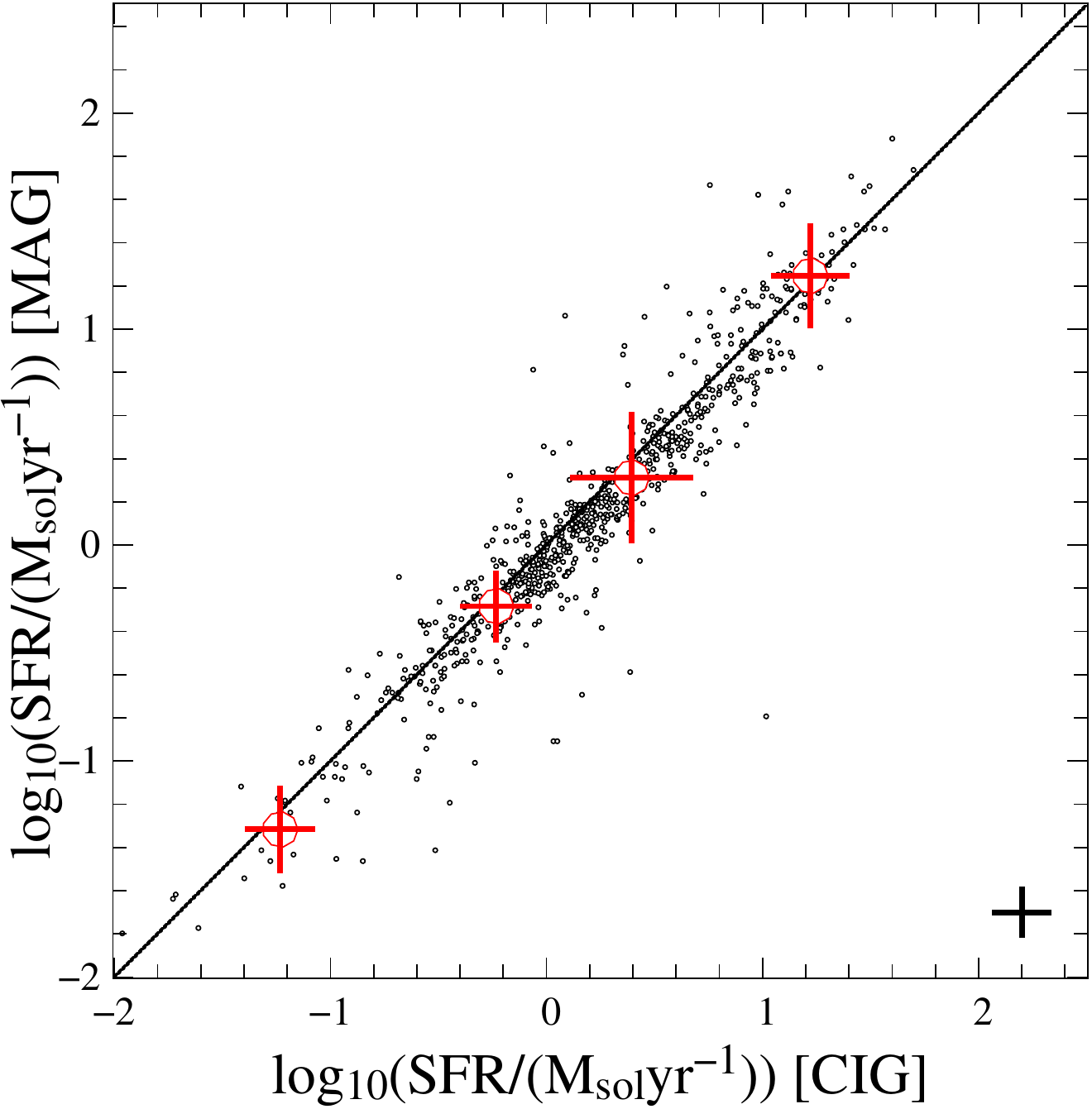}
\includegraphics[clip=,width=.4\textwidth]{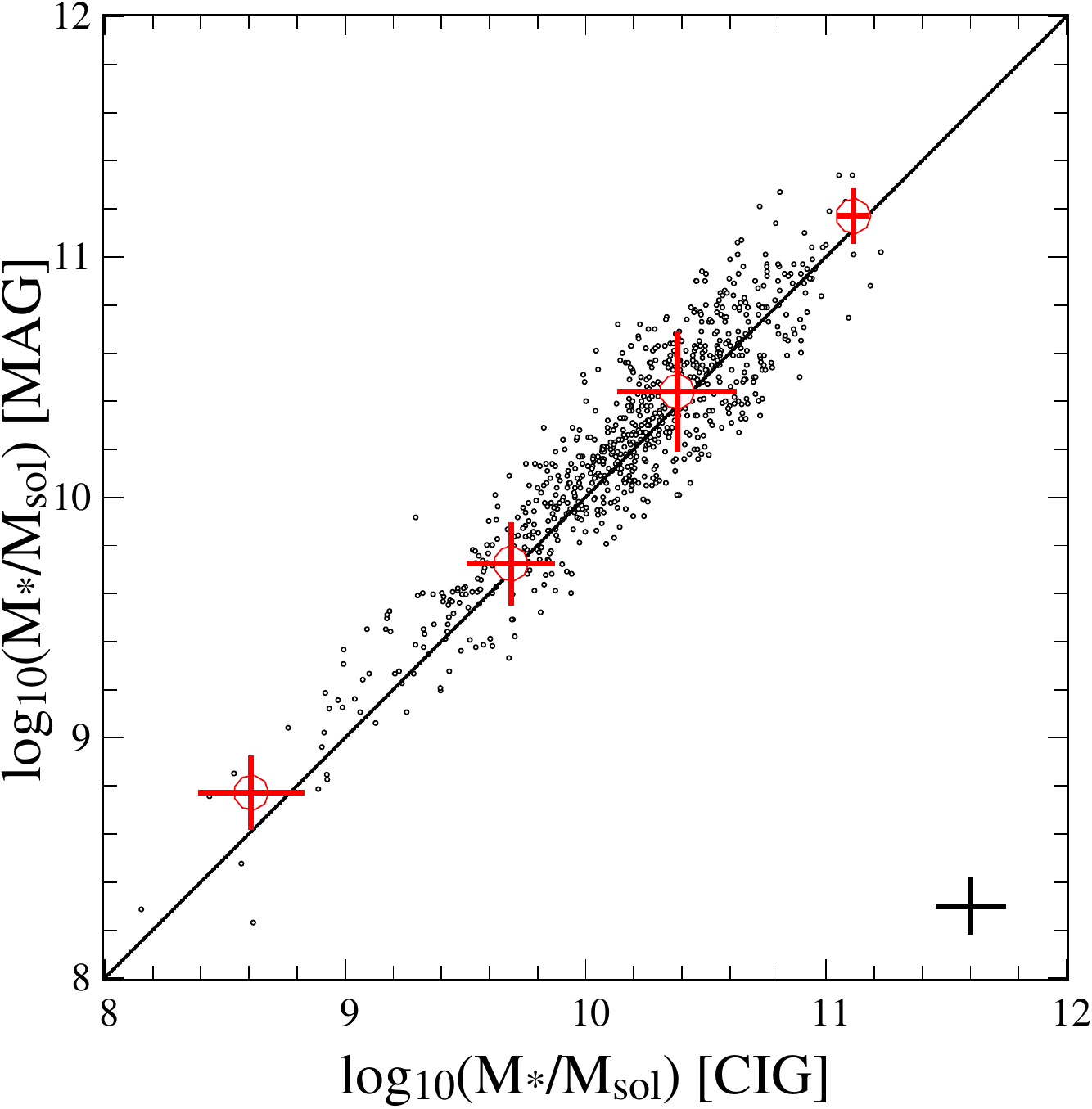}
\includegraphics[clip=,width=.4\textwidth]{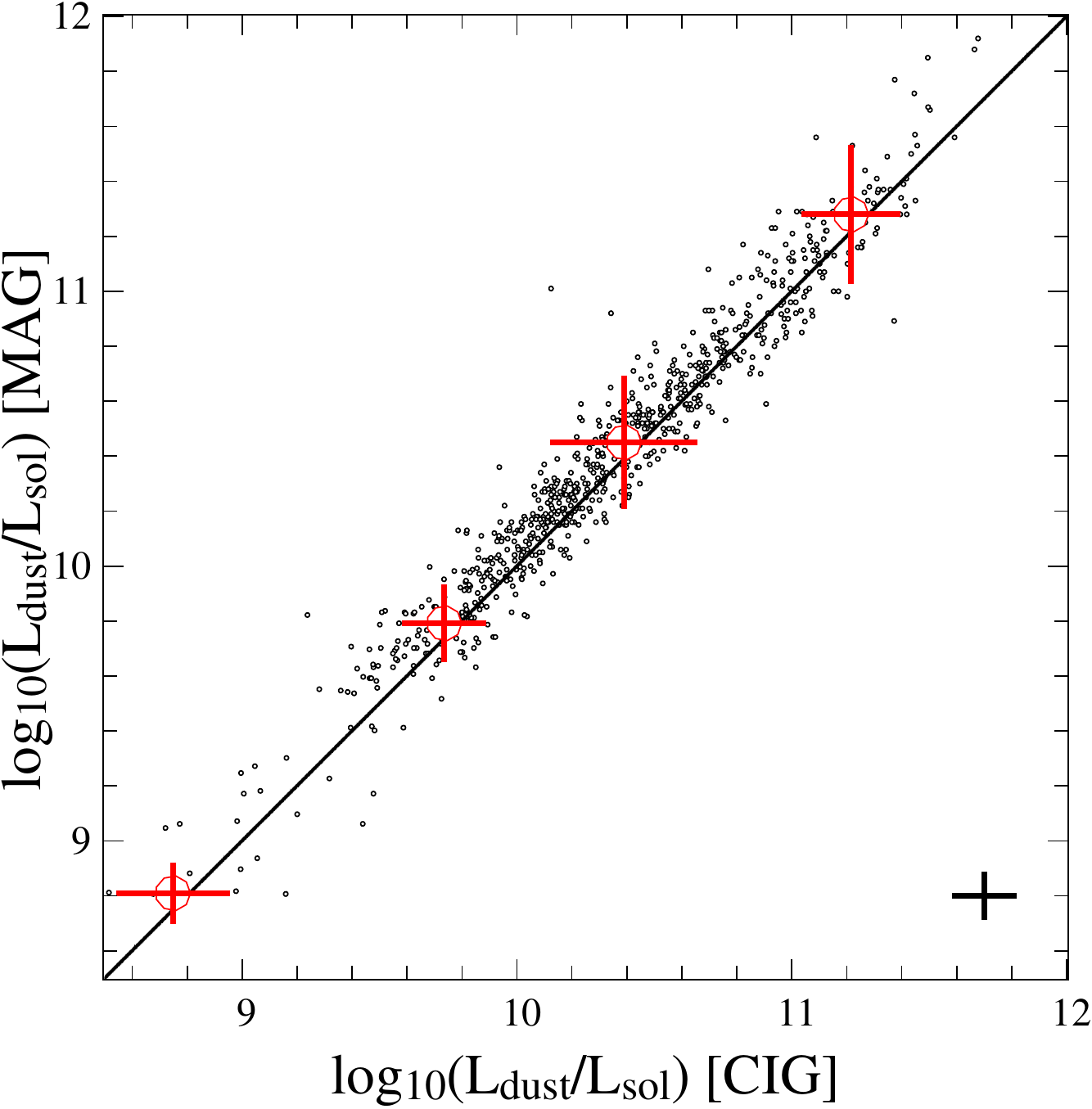}
\end{center}
\caption{SFR (top panel), $M_\ast$ (middle panel), and $L_{dust}$ (bottom panel) obtained with MAGPHYS (ordinate) and CIGALE (abscissa). Red circles show the median estimated inside bins of $\Delta log_{10}(x)$ = 1 with the associated variance. The solid line shows a linear relation between the two quantities. Average errors are shown in the bottom right corner.}
\label{comp}\end{figure}

\begin{figure}
\begin{center}
\includegraphics[clip=,width=.24\textwidth]{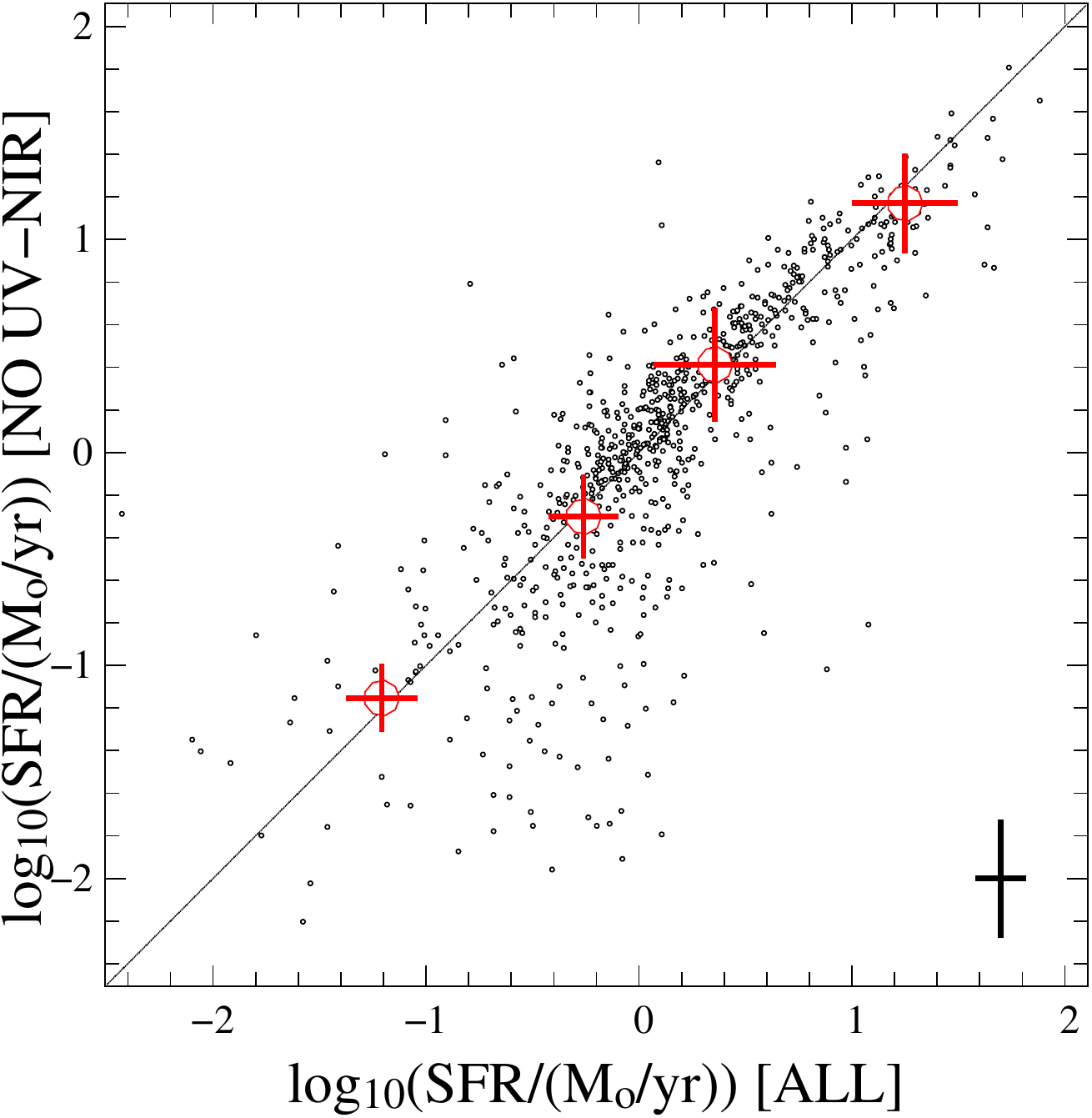}
\includegraphics[clip=,width=.24\textwidth]{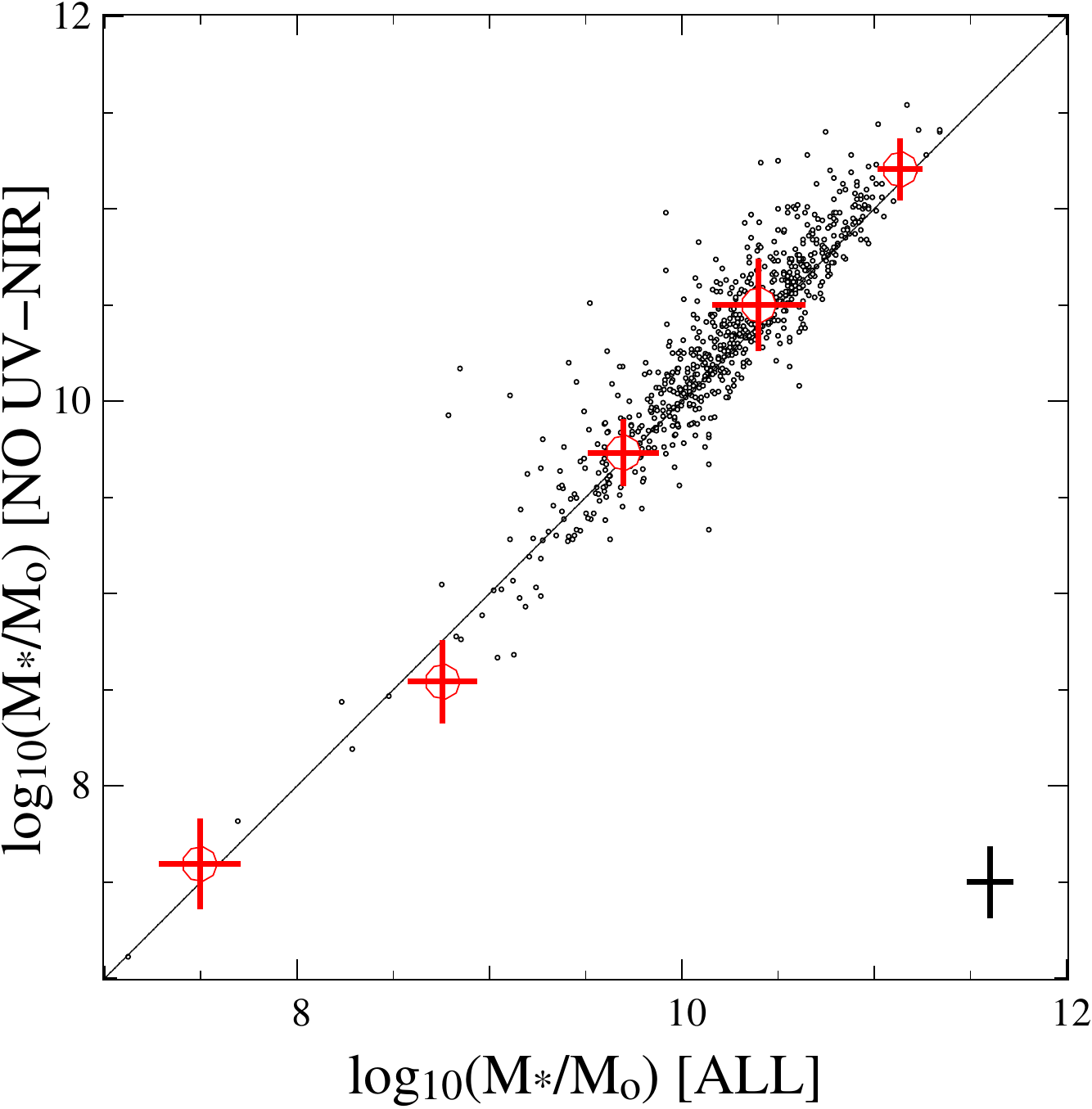}
\includegraphics[clip=,width=.24\textwidth]{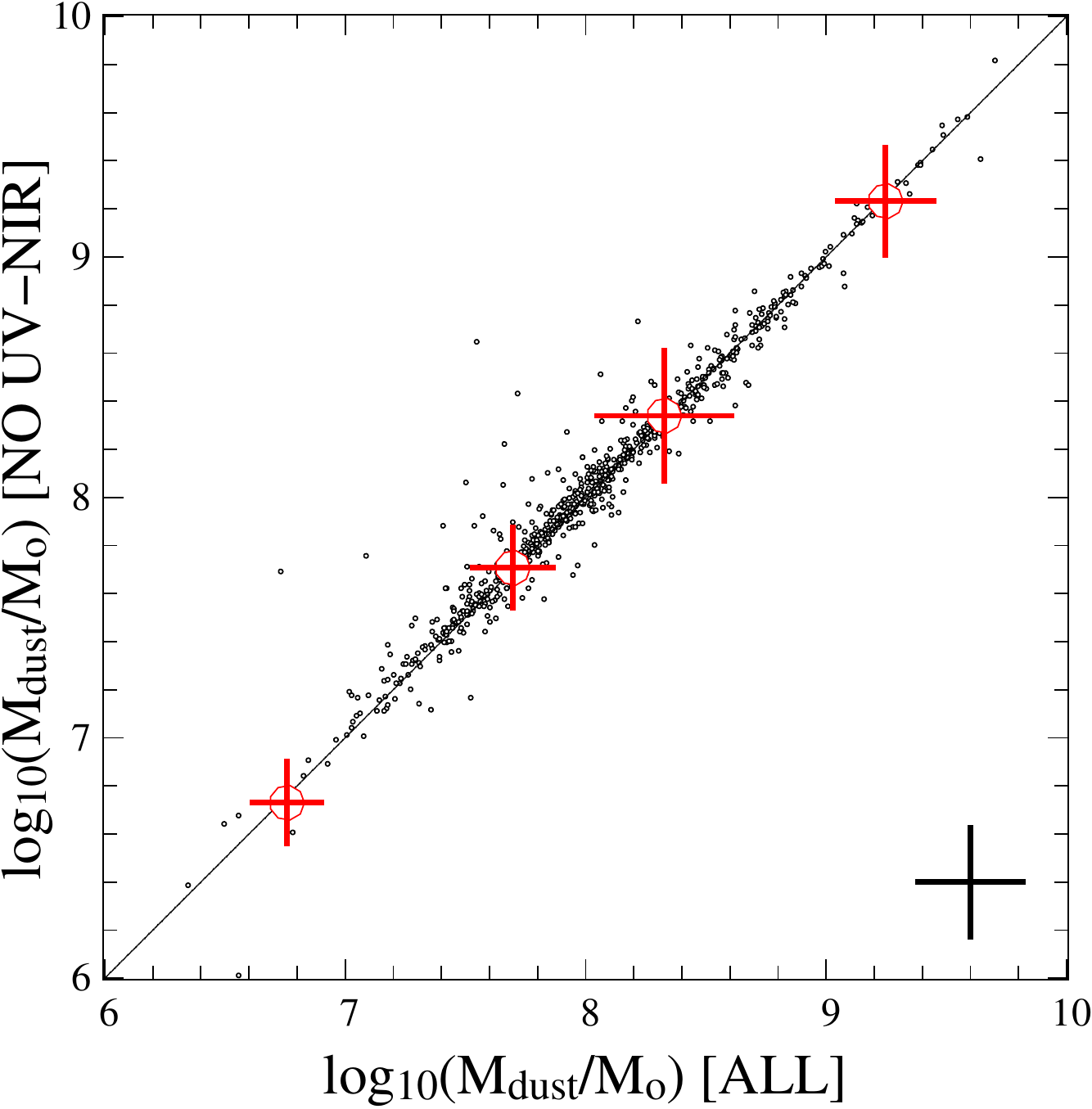}
\includegraphics[clip=,width=.24\textwidth]{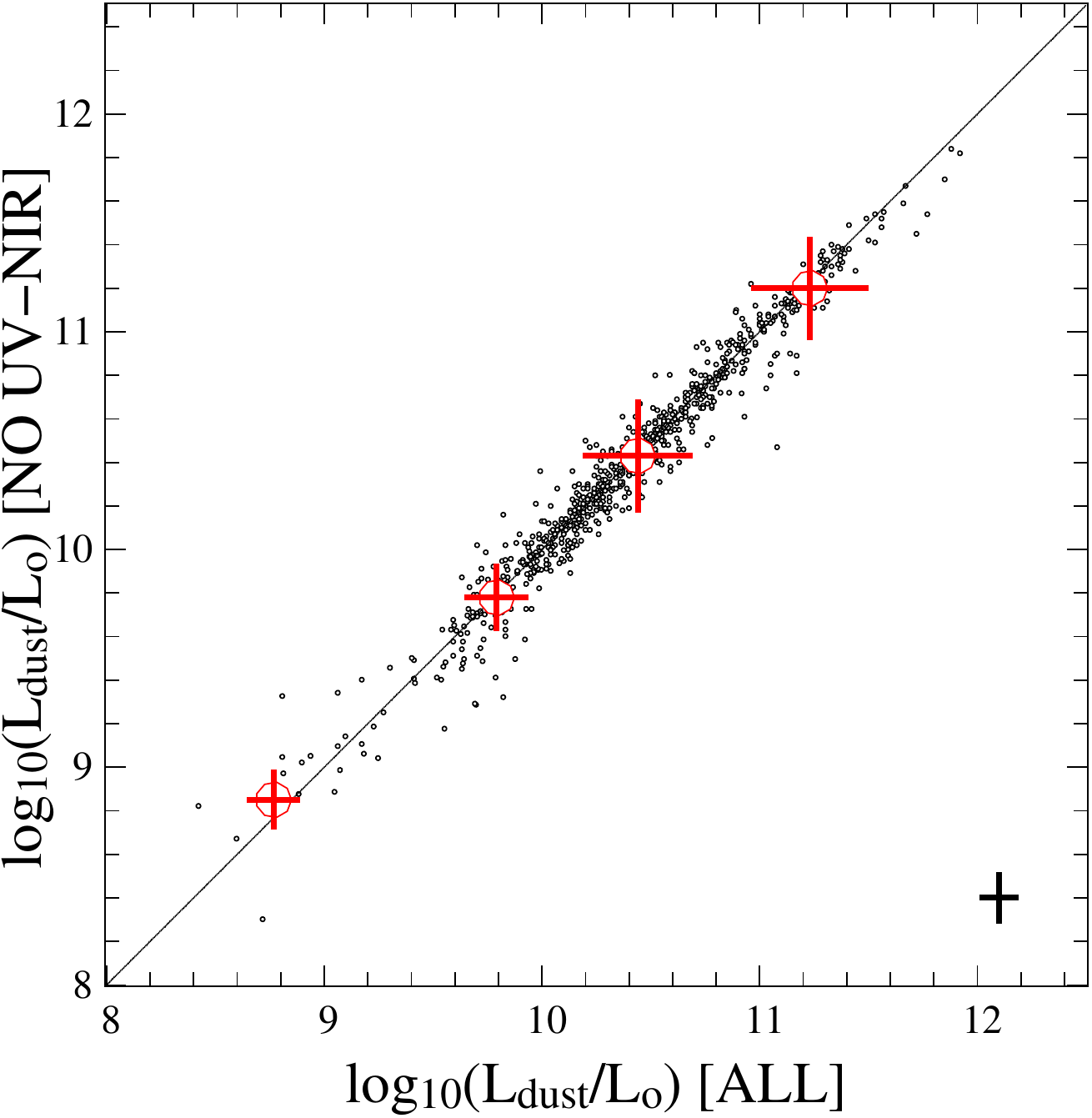}
\end{center} 
\caption{SFR (top left panel), $M_\ast$ (top right), $M_{dust}$ (bottom left), and $L_{dust}$ (bottom right panel) obtained from the main sample (abscissa) and the extended sample (ordinate). The dashed line shows a linear relation between the two quantities. Red circles show the median calculated inside bin evenly spaced with associated variance. Average errors are shown in the bottom right corner.}
\label{compext}
\end{figure}

\section{Method}
\label{met}

The broad-band panchromatic data set was fitted using two different SED fitting techniques: MAGPHYS\footnote{\tt http://www.iap.fr/magphys/magphys/MAGPHYS.html} \citep{dac2} and CIGALE\footnote{\tt http://cigale.lam.fr/} \citep{nol}. Both methods interpret a galaxy SED as a combination of different simple stellar population libraries and dust emission spectra. 

The first step of MAGPHYS is to build two different set of libraries that reproduce the stellar and dust emission of a galaxy: libraries containing attenuated emission from stellar populations, from 91\AA\ to 160 \mic, are built using the models of \cite{bru} with the \cite{cha} prescription for dust attenuation and a \cite{chab} initial mass function (IMF). 

The total dust emission from MIR to sub-mm bands is assumed to be the sum of a component due to the dust heated in the stellar birth clouds, and a component originating from the diffuse interstellar medium (ISM). Dust emission due to the radiation produced in the stellar birth cloud is the sum of polycyclic aromatic hydrocarbons (PAH), hot MIR continuum, and a warm component in thermal equilibrium with temperatures between 30-70 K. Emission originating from dust in the diffuse ISM is modeled as the sum of the three components mentioned above, with the addition of a cold component in thermal equilibrium with temperature between 10-30 K. This range of temperatures is slightly wider than the MAGPHYS version in \cite{dac2}: an extended library has been recalculated by E. da Cunha and tested in \cite{via} and \cite{agi}. Dust in thermal equilibrium is optically thin and emits as a modified black body, $\propto \kappa_\lambda B_\lambda(T)$, with the dust absorption coefficient modeled as $\kappa_\lambda \propto \lambda^{-\beta}$ and $\beta$ equals 1.5 and 2.0 for warm and cold components, respectively. For the stellar emission component, MAGPHYS includes 25000 spectra using different stochastic SFHs\footnote{Exponentially decreasing SFR with bursts randomly superposed}, metallicities, and dust attenuation, while dust emission is represented by 50000 spectra using different combinations of warm and cold dust, PAH, and MIR continuum \citep[see][for further details]{dac2}. The method requires an energy balance between the energy absorbed by dust in the UV-optical and the energy re-emitted in the form of IR radiation by dust itself. This allows us to constrain consistently the stellar and dust emission. The fraction of stellar radiation absorbed by dust in the stellar birth clouds and in the ambient ISM is redistributed at FIR wavelengths, assuming that the starlight is the only source of heating. With these assumptions we can then compare our multiwavelength data with a different combination of theoretical SEDs and calculate the $\chi^2$ parameter, the usual goodness of fit estimator. These values allow a probability density function (PDF) -- whose median corresponds to the best estimate of the true value -- to be built for each physical parameter \citep[see also][]{dac}.

CIGALE estimates the stellar component of a galaxy exploiting emission models that assume different IMF and stellar libraries. In our case we chose a \cite{chab} IMF and the SEDs of \cite{bru2} convolved with different exponentially decreasing SFHs whose e-folding $\tau$ are listed in Table \ref{cigparam}. Dust emission is represented by the \cite{dra} model, which reproduces the dust emission assuming a composition of amorphous silicate and carbonaceous grains with the size distribution of the Milky Way \citep{wei}. Dust can be heated by hot O-B stars in photo dissociation regions or by the large number of stars that produce diffuse radiation. The former in \cite{dra} models is parameterized with $U_{min}$, i.e., the intensity of the interstellar radiation field and the fraction of the total dust heated by these two components is given by $\gamma$, with values between 0 and 1. Another parameter introduced is the abundance of PAH ($q_{PAH}$) which quantifies the contribution of the PAH to the total dust emission. Defining a set of starting parameters (see Table \ref{cigparam}) CIGALE builds a library of SEDs from which is estimated the $\chi^2$.

\begin{table}\caption[]{Parameters used in the CIGALE fitting procedure.}\label{cigparam}
\begin{center}\begin{tabular}[0.8\textwidth]{cc}\hline \hline
Parameter  & Values\\ 
\hline
\\
$\tau$ [Gyr]& 0.25, 0.5, 1, 2, 3, 4, 5, 7, 9, 14\\
$U_{min}$  & 0.1, 0.15, 0.2, 0.25, 0.35, 0.5, 0.6, 0.7, 0.8, 1.0, 1.2, 1.5\\
$q_{PAH}$  & 0.47, 1.12, 1.77, 2.5, 3.19, 3.9, 4.58, 5.95\\
$\gamma$ & 0.001, 0.01, 0.1, 0.9\\
\\
\hline \hline
\end{tabular}\end{center}\end{table}

To determine for both methods the range of acceptable $\chi^2$ values, we considered the $\chi^2$ distribution in the main samples and we built the probability density function, shown as a black solid and red dashed line in Fig. \ref{chi2} for MAGPHYS and CIGALE, respectively. The theoretical probability density function of the $\chi^2$ distribution is

\begin{equation}
P\propto \frac{1}{2^{N_{dof}/2}\cdot \Gamma(N_{dof}/2)}(\chi^2)^{(N_{dof}/2)-1} \ {\rm exp}^{-\chi^2/2},
\label{chif}
\end{equation}
where $\Gamma(x)$ represents the Gamma function, and $N_{dof}$ is the number of degrees of freedom of the problem. The degrees of freedom of our system are not simply the number of photometric points ($N_{bnd}$) minus one because neighboring photometric bands are not independent. For this reason we fit the PDF in Fig. \ref{chi2} using -- in Eq. \ref{chif} -- the relation between $N_{dof}$ and $N_{bnd}$ determined in \cite{smi}: 

\begin{equation}
N_{dof} \approx -2.82+0.66 \cdot N_{bnd}+7.91 \cdot 10^{-3} \cdot N_{bnd}^2.
\label{ndof}
\end{equation}

The best fits shown as a black dotted and red dot-dashed line in Fig. \ref{chi2} allow the 99\% confidence interval of the $\chi^2$ distribution to be determined, indicating the threshold above which there is only a 1\% probability that our SED fitting is reliable. We found a median value $\chi^2$ of 2.6$_{-1.2}^{+6.3}$ and 1.9$_{-0.9}^{+3.9}$ (16th and 84th percentiles) for MAGPHYS and CIGALE with less than 1\% of the sources outside the range defined by Eq. \ref{chif}. On average, MAGPHYS recovers higher $\chi^2$ values than CIGALE, and this could indicate a higher accuracy in the fit process for the latter. However, the best fit obtained using Eq. \ref{chif} tells us that most of the fits are equally acceptable once they are in the 99\% percentile defined by Eq. \ref{chif}. Because of the number of degrees of freedom of the problem is not well defined and can only be approximated by Eq. \ref{ndof}, the issue here is that a normalized $\chi^2$ close to 1 does not necessarily indicate a good fit. For this reason we still keep both methods for the analysis, quantifying differences and analogies in Sect. \ref{res}.

\section{Results}
\label{res}

Before proceeding with the analysis, we compare the results obtained with MAGPHYS and CIGALE, restricting the analysis to the parameters investigated in this paper and leaving a more extended study for a future paper (Pappalardo et al. in prep.).

\subsection{Comparison of MAGPHYS and CIGALE results}

Fig. \ref{comp} shows the relation between SFR, $M_\ast$, and dust luminosity ($L_{dust}$) estimated in MAGPHYS (ordinate) and CIGALE (abscissa). The SFR, $M_\ast$, and $L_{dust}$ are consistent in both methods, with variations below 20\%. We divided each parameter in different bins and we estimated the median, shown as a red circle in Fig. \ref{comp} with the associated variance. Both methods produce results consistent with a linear relation (solid line), implying a substantial equivalence for the parameters considered. From now on, unless expressly written, we refer to the results of MAGPHYS.

We also quantified how the results vary when removing the constraints at UV and NIR wavelengths, as done for the extended sample in Sect. \ref{hd}. Fig. \ref{compext} compares SFR, $M_\ast$, $L_{dust}$, and the dust mass ($M_{dust}$) obtained with MAGPHYS considering only the galaxies in common. The four quantities are consistently calculated in both the main and the extended sample. The main differences are found in the SFR, with an increasing scatter at low values. However, they are mostly limited to galaxies with moderate star formation, SFR $<$ 1 M$_\odot$ yr$^{-1}$, a range where its determination is in any case uncertain. We also note that stellar masses are consistent in both samples. This is a strong indication that the combination of WISE and optical SDSS data is sufficient to constrain the fit in this spectral region, and hence the stellar mass value derived. In conclusion, removing the UV and the NIR from the analysis still results in reliable fits, once an extended photometric coverage all over the SED is provided. We can then reasonably assume that the results obtained in the extended samples are reliable and robust with respect to the quantities analyzed.

\begin{figure} 
\begin{center} 
\includegraphics[clip=,width=.24\textwidth]{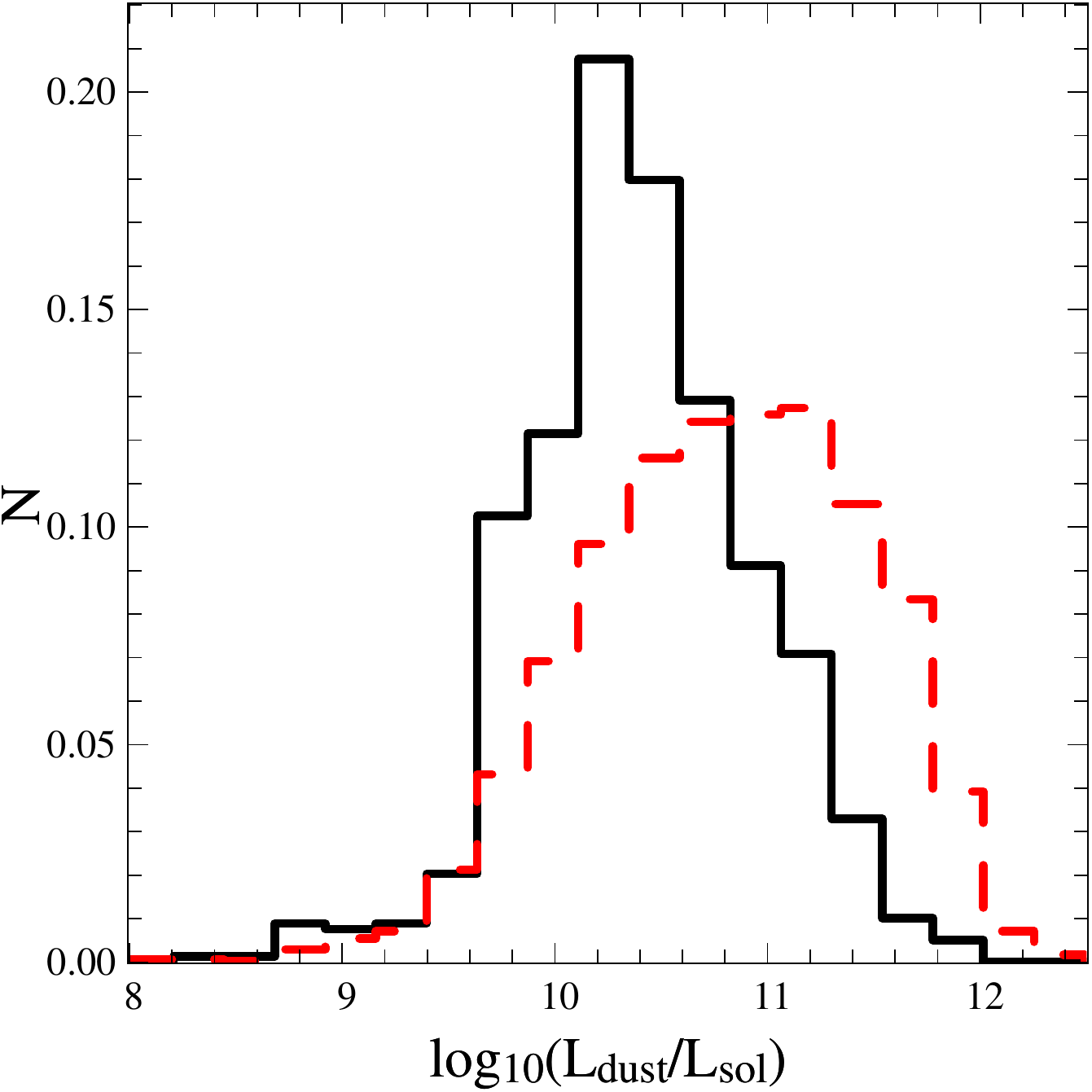}
\includegraphics[clip=,width=.24\textwidth]{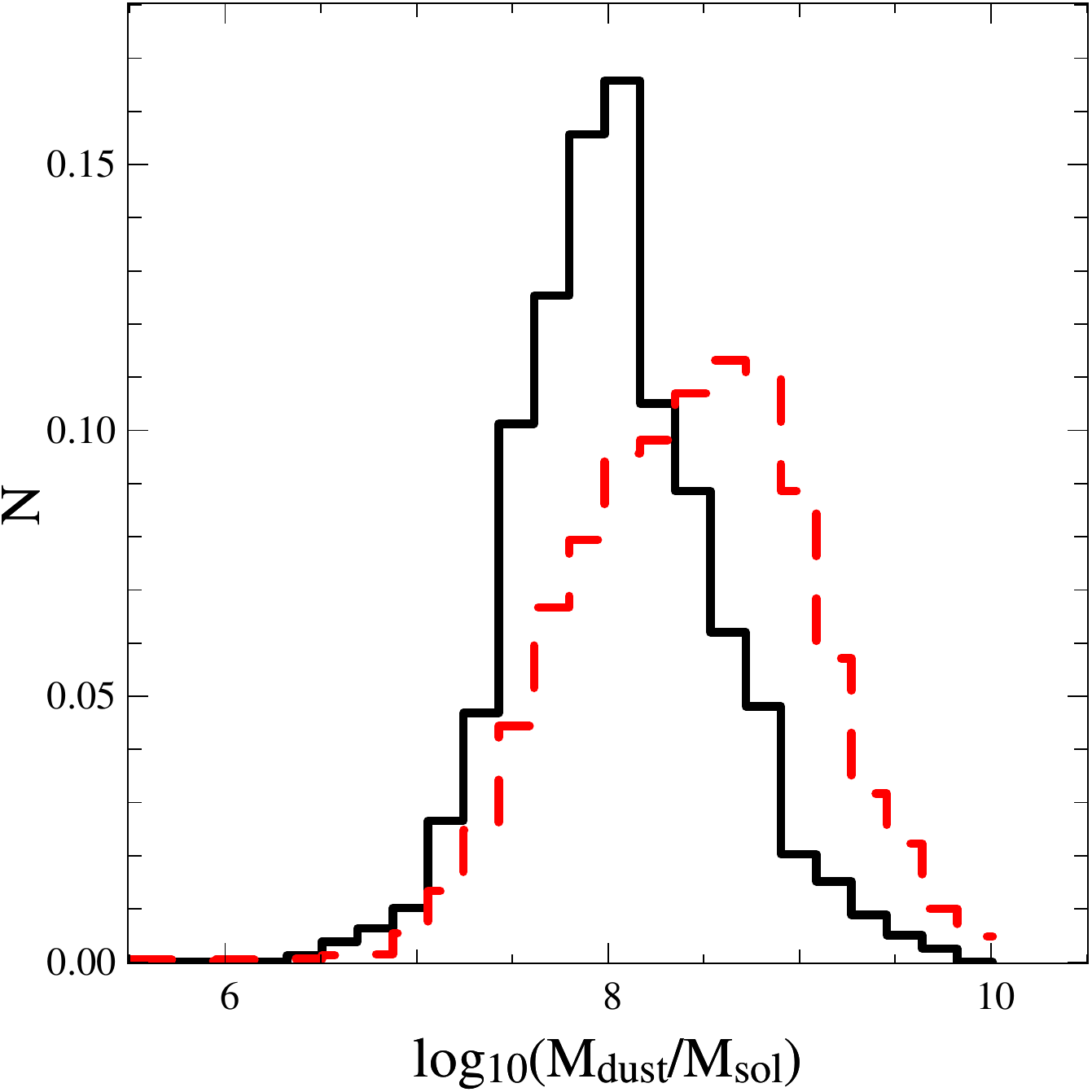}
\includegraphics[clip=,width=.24\textwidth]{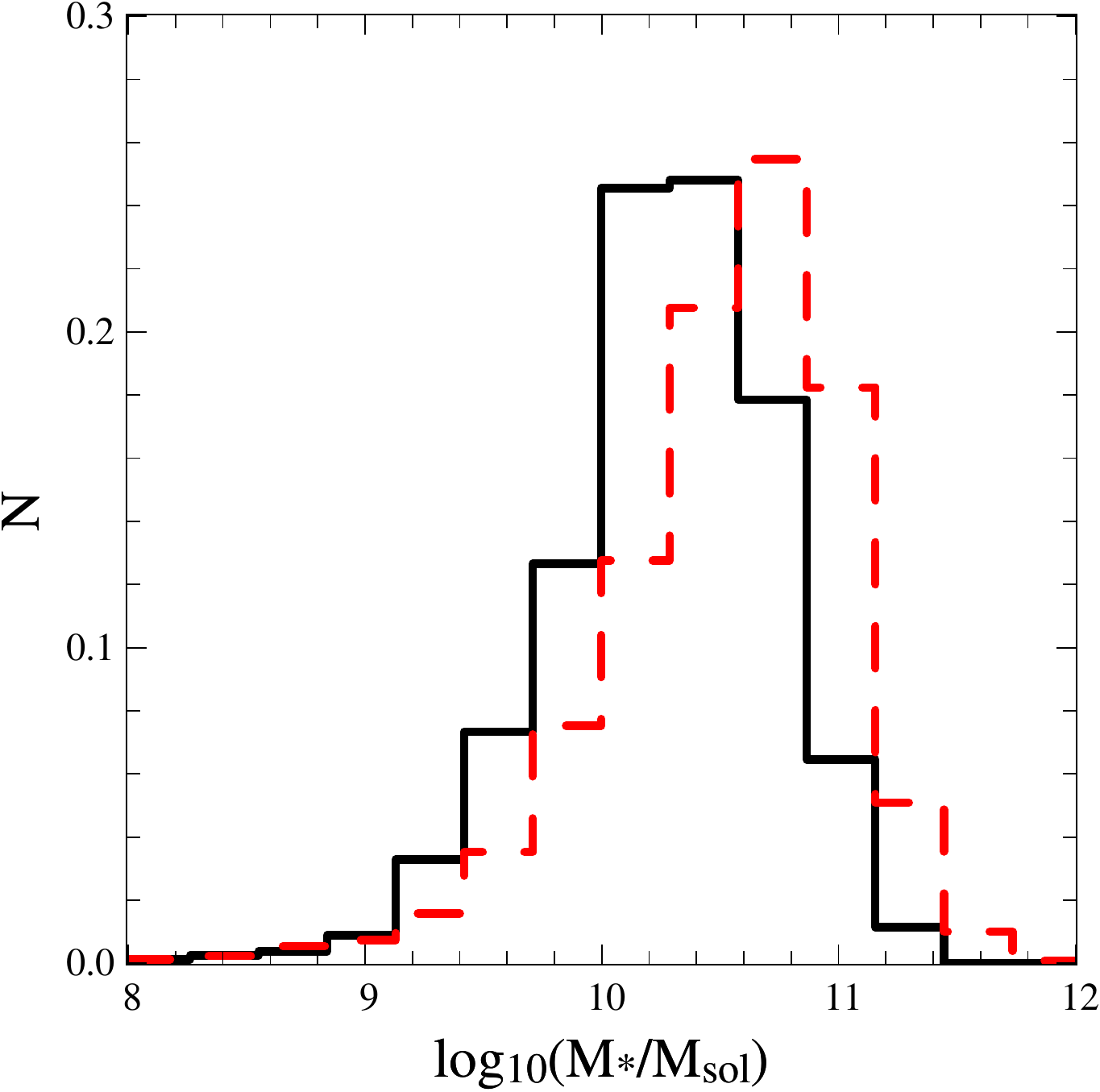}
\includegraphics[clip=,width=.24\textwidth]{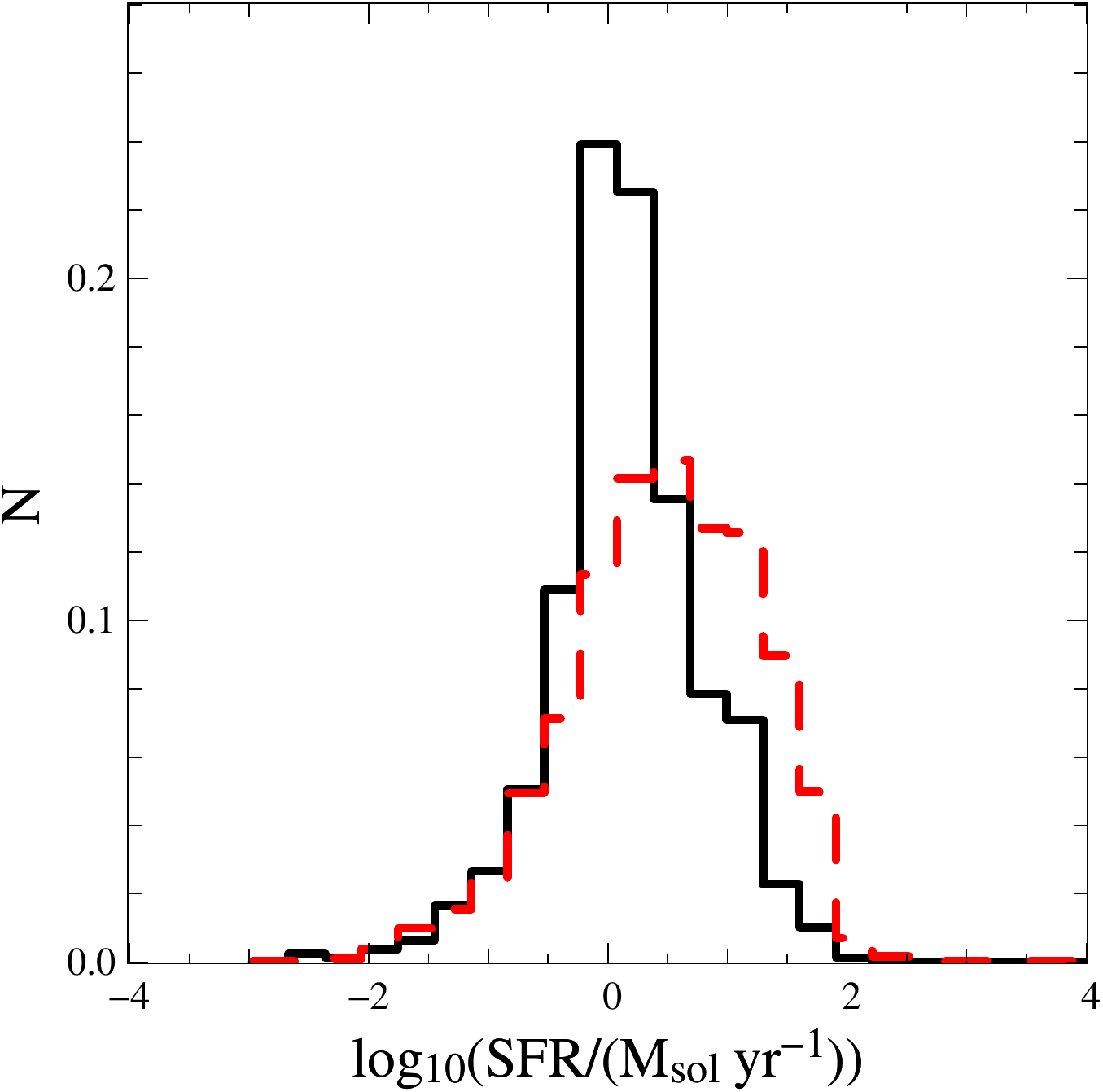}
\end{center} 
\caption{Normalized distribution of dust luminosity (top left), dust mass (top right), stellar mass (bottom left), and SFR (bottom right) for the main (black solid line) and extended (red dashed line) sample, estimated with MAGPHYS.}
\label{his} 
\end{figure}

\subsection{Properties of low FIR luminosity sample}
\label{pfir}

\begin{table*}
\caption{\label{t7}Estimated physical parameters.}
\label{tabella}
\begin{center}
\begin{tabular}[0.8\textwidth]{ccccc}
\hline \hline
Parameter  & Main Sample & Extended sample & H-ATLAS & SINGS\\
(1) & (2) & (3) & (4) & (5)\\ 
\hline\\
$M_{dust}$ [M$_\odot$] & 9.8$\pm$1.6 $\times$ 10$^7$ & 2.6$\pm$1.4 $\times$ 10$^8$ & 5.5 $\times$ 10$^7$ & 1 $\times$ 10$^8$\\
$L_{dust}$ [L$_\odot$] & 2.3$\pm$0.3 $\times$ 10$^{10}$ & 6.2$\pm$2 $\times$ 10$^{10}$ & 6.4 $\times$ 10$^{10}$ & 6.4 $\times$ 10$^{10}$\\
SFR [M$_\odot$ yr${-1}$] & 1.2$\pm$0.3 & 2.8$\pm$4.7 & 3.25 & 4.17\\
$M_\ast$ [M$_\odot$] & 1.9$\pm$0.1 $\times$ 10$^{10}$ & 3.5$\pm$0.1 $\times$ 10$^{10}$ & 3 $\times$ 10$^{10}$ & 2.5 $\times$ 10$^{10}$\\
$M_{dust}/M_\ast$ & 0.018$\pm$0.002 & 0.025$\pm$0.001 & 0.002 & 0.004\\
 \hline \hline \end{tabular}\end{center}
 \tablefoot{Median obtained for different physical parameters (Col. 1) estimated with MAGPHYS in the main sample (Col. 2) and the extended sample (Col. 3) with associated errors estimated as in \cite{row}. Columns 4 and 5 show median values from \cite{smi} and \cite{dac}, respectively.}\end{table*}

\begin{figure}\begin{center}
\includegraphics[clip=,width= .49\textwidth]{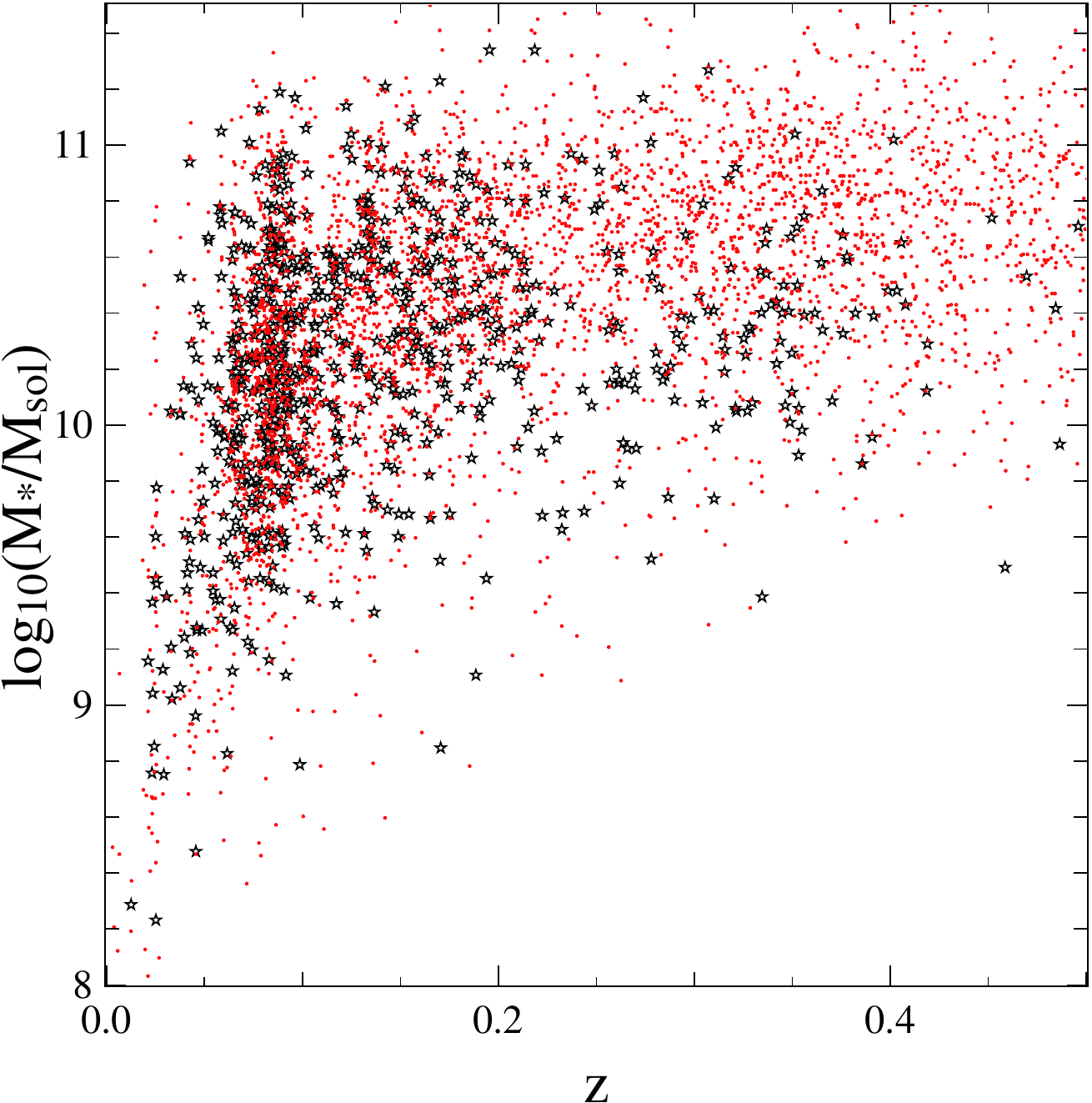}
\end{center}
\caption{Stellar mass as a function of redshift for the main (black stars) and the extended (red dots) sample.}
\label{msz}\end{figure}

Figure \ref{his} shows the normalized distribution of $M_{dust}$, $L_{dust}$, $M_\ast$, and SFR for the main and the extended sample. Removing the constraints that the SED should have UV data points, we include galaxies with higher dust extinction and, hence, a larger dust content. This explains why, in Fig.~\ref{his}, the normalized distribution peaks at higher values of the dust mass and luminosity. The fact that the SFR peaks at slightly higher values as well is most likely an indirect effect, coming from the correlation between dust content and star formation rate. The small difference in the peak of the stellar mass distribution is due to the fact that in the extended sample there are more distant galaxies (see Fig.~\ref{red}), and we are selecting slightly more massive ones. This is confirmed in Fig. \ref{msz}, where we show the stellar mass as a function of redshift for the main and the extended sample. Despite a different number of objects at $z > 0.2$, we see that most of the galaxy populations explored in both samples uniformly span a range of stellar masses between 10$^9 <$ M$_\ast <$ 10$^{11}$ M$_\odot$ at redshift $z \sim$ 0.1. Table \ref{tabella} shows the median values of the main physical parameters in our samples compared to the H-ATLAS works of \cite{smi} and the {\it Spitzer} 24 $\mu$m selected sample of \cite{dac}. Both studies investigated low redshift galaxies and used MAGPHYS. Our main sample has a median $L_{dust}$ about three times lower than the other two samples, and a median stellar mass two times lower than H-ATLAS. These values justify the choice of the HeViCS survey in order to tackle the galaxy population with low mass and low dust luminosity at $z <$ 0.5. 

\begin{figure*}\begin{center}
\includegraphics[clip=,width= .79\textwidth]{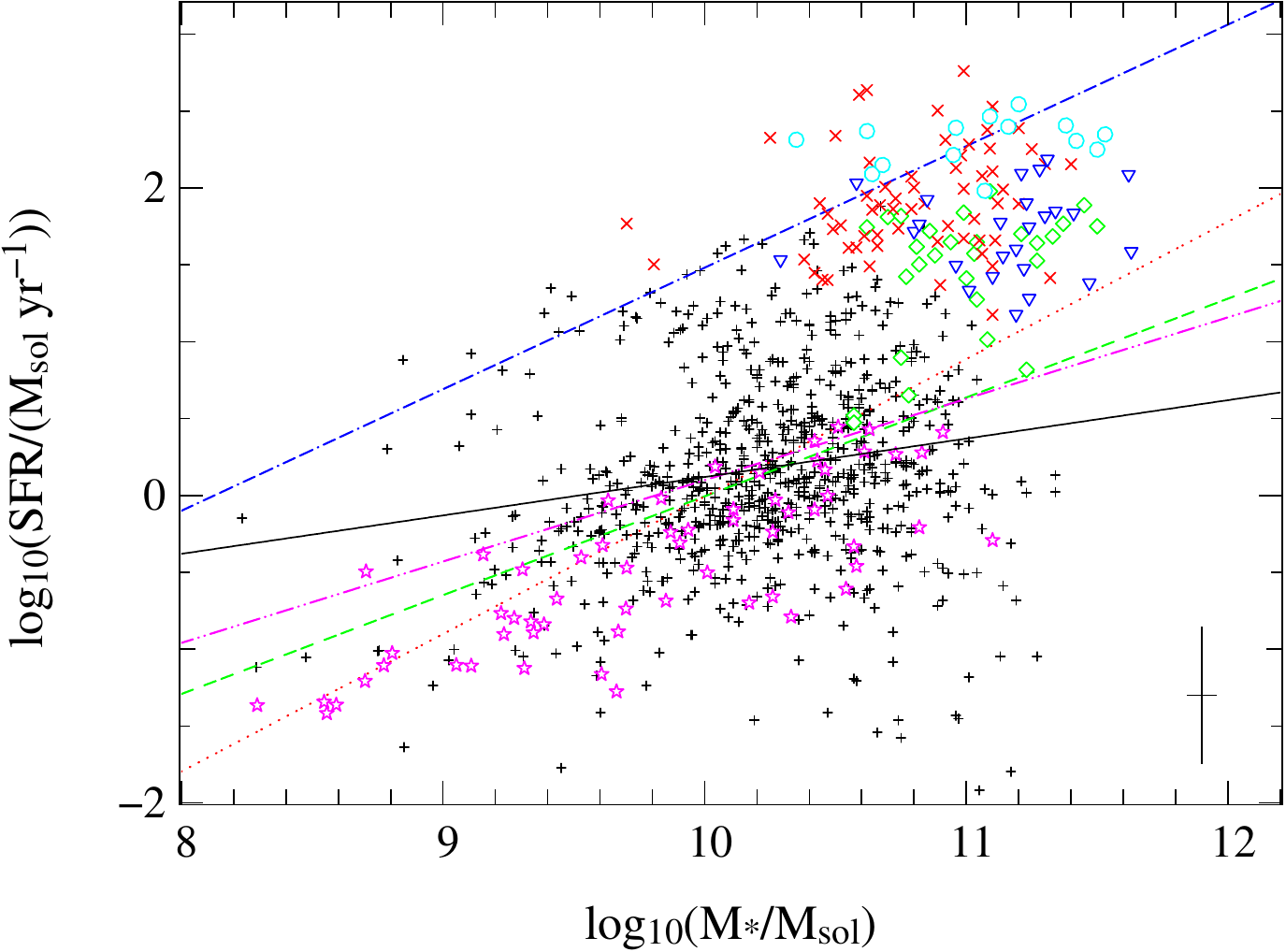}
\end{center}
\caption{SFR-$M_\ast$ relation in the main sample (black crosses) with average error bars shown in the bottom right corner and best fit estimate (black solid line). Green diamonds show the results for a sample of low redshift star forming galaxies \citep[0.05 $< z <$ 0.5,][]{bau}, red crosses are high redshift star forming galaxies \citep[1 $< z <$ 3][]{tac}, blue triangles local LIRGs at $z \approx$ 0.4 \citep{gea}, magenta stars and cyan circles show local spirals \citep{ler} and ULIRGs \citep{gao}, respectively. The fits are from: local galaxies of \cite{cie} and \cite{pen2} (green dashed and red dotted line), galaxies at $z = 2$ and $z = 4$ in \cite{rod} (blue dot-dashed line), and the compilation shown in \cite{spe} (magenta double dotted line).}
\label{mseq}\end{figure*}

Other large surveys, e.g., those cited in Sect.~\ref{intro}, which include observations with \hers, showed some limitations from this point of view: for example, \cite{lee} cross-correlate data from HerMES and COSMOS, and investigate the relation between SFR and $M_\ast$ in a sample of $\sim$4000 sources. Surprisingly their population does not follow the nominal main sequence, but appears to have a much flatter distribution. They explain this difference saying that ``{\it Herschel} observations of COSMOS generally sample only the most luminous regime of the SFR/$M_\ast$ plane, so we cannot make any statements about galaxies at lower luminosities''. On the other hand, for H-ATLAS \cite{smi} showed that with only two cross scans they select mainly galaxies with dust luminosities typical of luminous infrared galaxies (L$_{FIR} > 10^{10}$ L$_\odot$, LIRGs).

\begin{figure*}\begin{center}
\includegraphics[clip=,width= .79\textwidth]{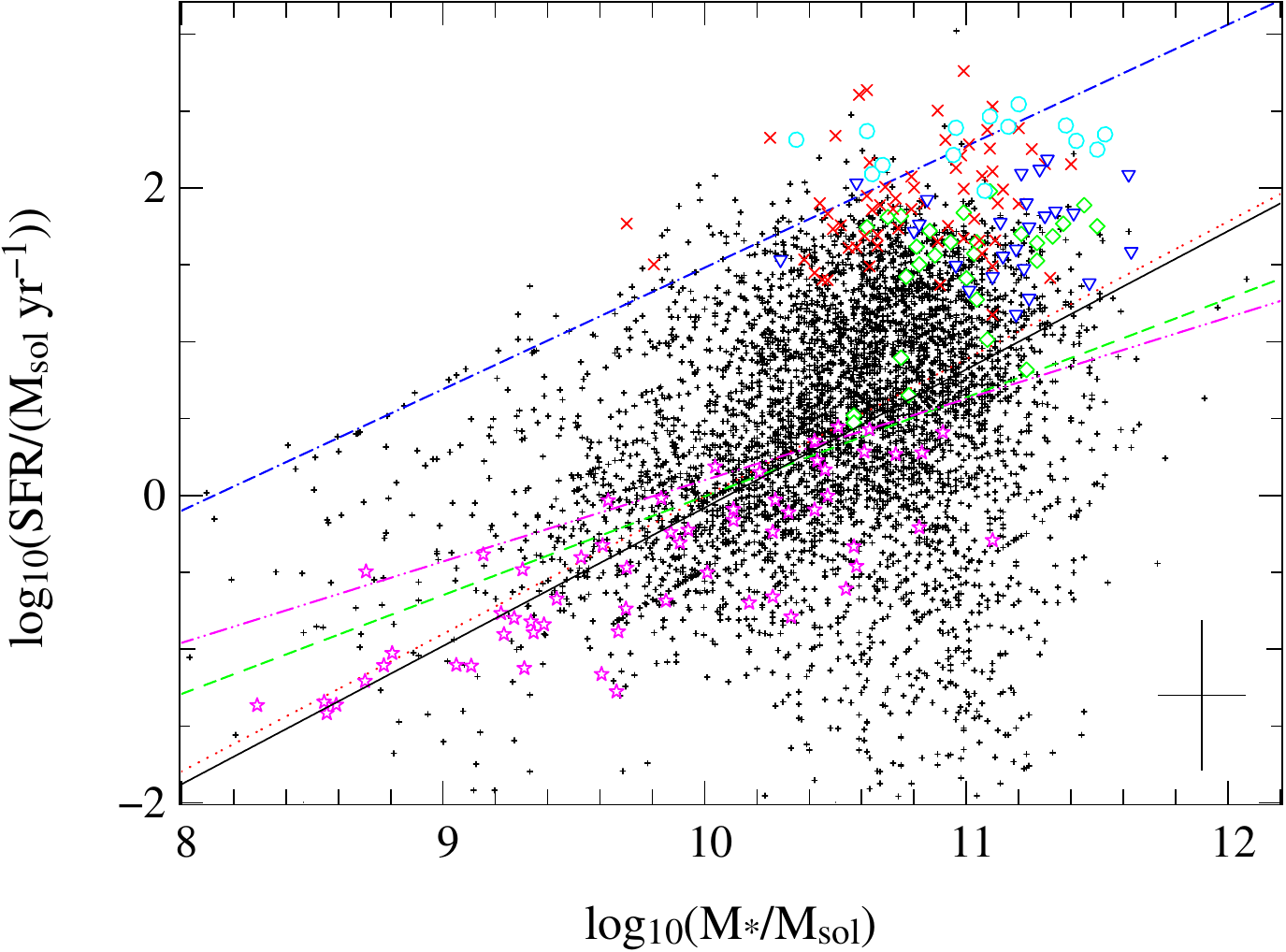}
\end{center}
\caption{SFR-$M_\ast$ relation in the extended sample (black crosses) with average error bars shown in the bottom right corner. Symbols and lines are detailed in Fig. \ref{mseq}.}
\label{mseq2}\end{figure*}

Figures \ref{mseq} and \ref{mseq2} show the SFR-$M_\ast$ for the main and extended sample, together with the results for other samples from the literature. It is well known that most of star forming galaxies show a tight correlation in the SFR-$M_\ast$ plane called the main sequence \citep{pen,mar,mad2,alm}. In the local Universe this relation has a fixed slope between 10$^8<M_\ast<$10$^{11}$ M$_\odot$, while at higher redshift it offsets towards higher SFR, because of the larger gas supply. In Fig. \ref{mseq} the galaxies of the main sample are compared with a different data set representative of different galaxy evolution stages (see caption for details). The bulk of the main sample is formed of galaxies with moderate star formation comparable to local spirals (SFR $\approx$ 1 M$_\odot$ yr$^{-1}$) and $M_\ast\sim 10^{10}$ M$_\odot$. In addition to this component we observe at log$_{10}$(SFR/(M$_\odot$ yr$^{-1}$)) $>$ 0.5 a population of galaxies with similar stellar masses, but higher star formation, in a region of the SFR-$M_\ast$ plane intermediate between the local spirals (magenta stars) and higher redshift star forming galaxies (red crosses). These galaxies represent the population that at $z <$ 0.5 quenches its star formation activity and reduces its contribution to the total SFRD, as shown in \cite{hea}. 

However, most of this population is still not detected because in galaxies with $M_{dust} \sim$ 10$^8$ M$_\odot$ and SFR below 10 M$_\odot$ yr$^{-1}$ (Tab. \ref{tabella}), the weak UV radiation field produced during early star formation stages can be completely absorbed by the dust. This is the reason why we have defined an extended sample (see Sect.~\ref{hd}), where we also include dusty galaxies having lower UV luminosities. The main components of this population are galaxies with 10$^{10} <$ $M_\ast <$ 10$^{11}$ M$_\odot$ and SFR $\approx$ 10 M$_\odot$ yr$^{-1}$, shown in Fig. \ref{mseq2}.

Fig. \ref{mseq} and \ref{mseq2} identify two different aspects of galaxy evolution at $z <$ 0.5: most of the galaxies with $M_\ast >$ 3 $\times$ 10$^{10}$ M$_\odot$ are not in the main sequence, but are slightly offset, as a consequence of the decreased star formation. Low mass galaxies ($M_\ast <$ 1 $\times$ 10$^{10}$ M$_\odot$) do not undergo the same process; they are mostly settled in the main sequence with SFR and stellar masses consistent with local spirals. 

\begin{figure*} 
\begin{center} 
\includegraphics[clip=,width=.49\textwidth]{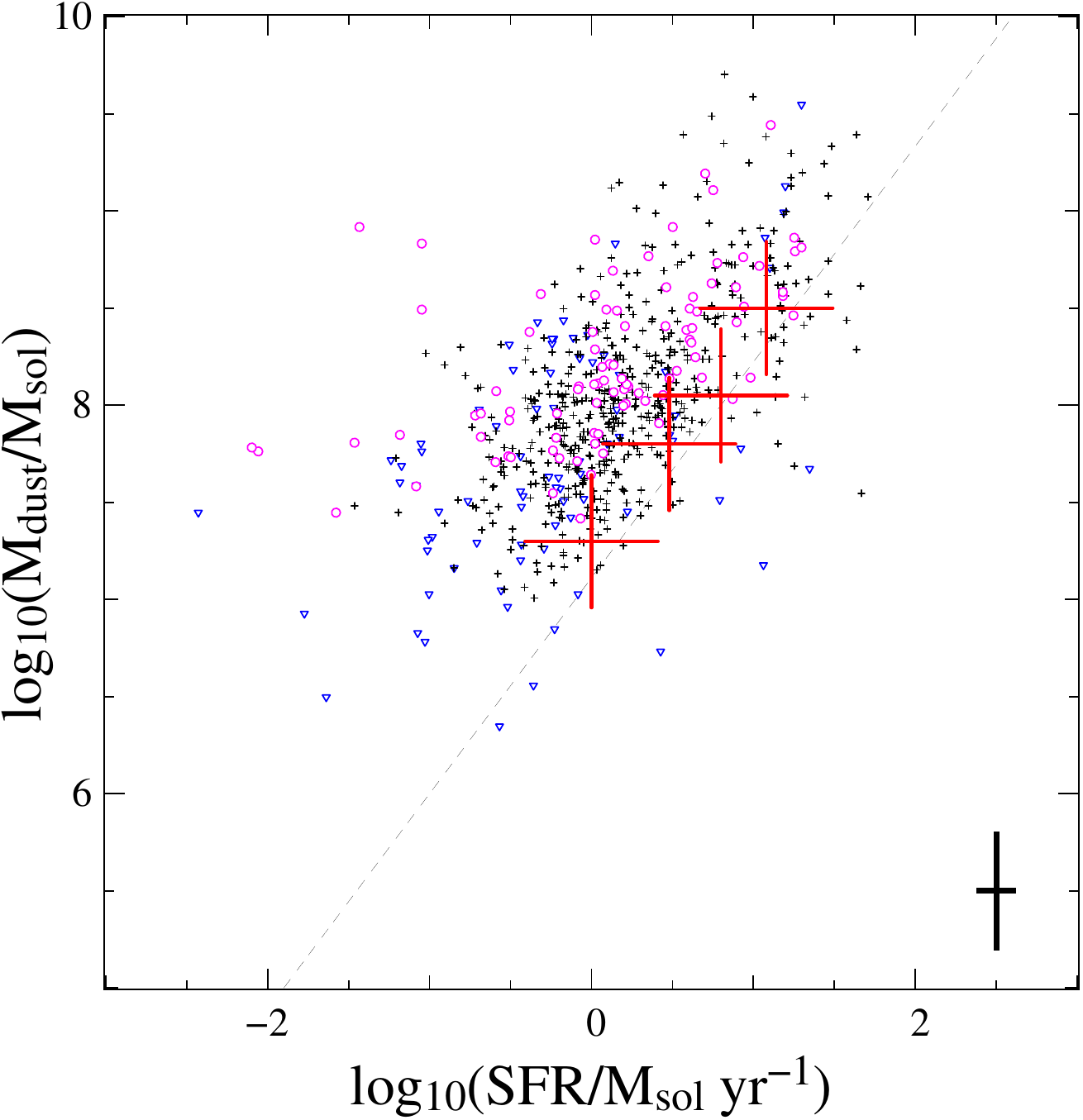}
\includegraphics[clip=,width=.49\textwidth]{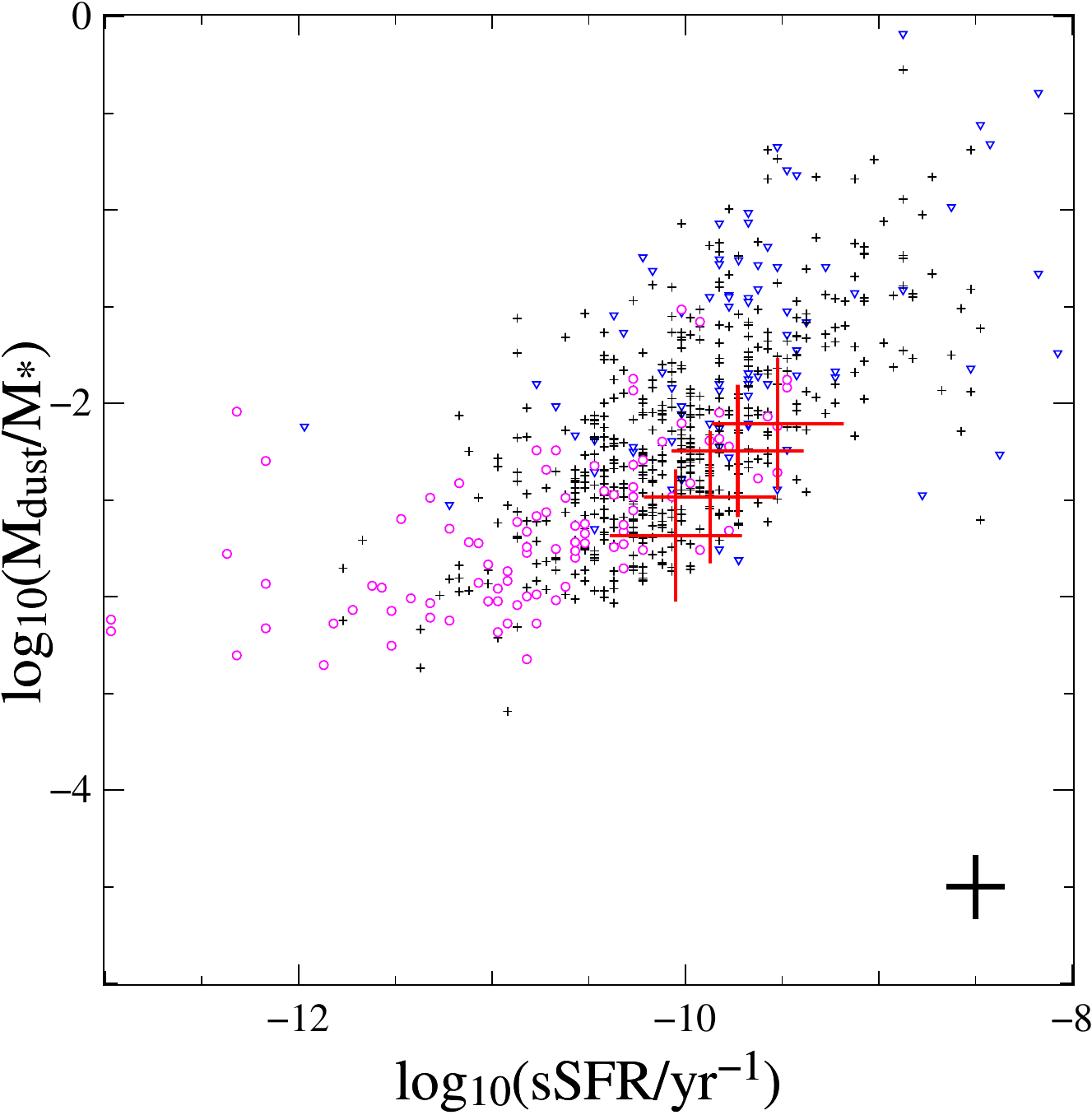}
\end{center} 
\caption{Dust mass (M$_{dust}$) vs SFR for the main sample. Red crosses show the results of \cite{smi} for a sample of $\sim$ 1400 objects selected at 250 $\mu$m from the H-ATLAS survey \citep{eal} at $z < 0.5$. The dashed line shows the best fit obtained in \cite{dac} from a sample of about $\sim$ 1700 low redshift galaxies. Blue triangles and the magenta circles show two subsamples of approximately the same size with $M_\ast <$ 5 $\times$ 10$^9$ M$_\odot$ and $M_\ast >$ 5 $\times$ 10$^{10}$ M$_\odot$, respectively. The cross in the bottom right corner shows the average error.}
\label{mdsfr} 
\end{figure*}

The previous analysis confirms that SFR estimates based on the FIR luminosity alone, should be used with caution. Different authors have already dealt with this problem: in normal galaxies\footnote{In this context ``normal galaxies'' means galaxies that are not undergoing a strong starburst.} dust can be heated not only by young stars forming clouds via UV radiation, but also by the diffuse interstellar radiation field produced by old stars during their late evolutionary stages \citep{ben2,ben3,ben4,boq,boq2,gro,via,cie,bos2,bos4,del2,del,smi3,smi4}. To investigate this aspect we consider two different relations: the dust mass vs the SFR, to find indications about the overall quantity of dust produced with respect to the current star formation, and the specific dust mass, i.e., $M_{dust}$/$M_\ast$ vs the specific SFR (sSFR = SFR/$M_\ast$), a parameter giving the same information as $M_{dust}$-SFR, but weighted with the actual mass of the galaxies. This is not a trivial aspect because high mass objects evolve more rapidly than low mass ones \cite[e.g.][]{ilb,gav}. In Fig. \ref{mdsfr} we show the results for the main sample. Red crosses reproduce the results of \cite{smi} for a sample of $\sim$ 1400 objects selected at 250 $\mu$m in the H-ATLAS survey \citep{eal} at redshift $z < 0.5$. In both samples the range of $M_{dust}$ is comparable, but the SFRs of our sources are systematically lower than galaxies analyzed in \cite{smi}. The average redshift of the H-ATLAS sample is $z\sim0.35$, with a difference $\Delta z \sim$ 0.2 with respect to our sources. This means that we are investigating a sample of galaxies at a slightly different stage of their evolution with respect to \cite{smi}, a stage in which the star formation activity is reduced. This is in agreement with many pieces of evidence which show that since $z\sim$ 1 the star formation density of the Universe has drastically decreased \citep{mad,hop,beh}. 

The dashed line in left panel of Fig. \ref{mdsfr} shows the best fit obtained in \cite{dac} from a sample of about $\sim$ 1700 galaxies with available GALEX, SDSS, 2MASS, and IRAS data. Galaxies have been selected with IRAS with a median dust mass of $M_{dust} \sim$ 5.5 $\times$ 10$^7$ M$_\odot$, which is slightly smaller than our value, $M_{dust} \sim$ 9.8$\pm$1.6 $\times$ 10$^7$ M$_\odot$. This difference in dust masses is due to a combination of two factors: the average redshift of the \cite{dac} sample is $z\sim$ 0.05, a range that is smaller than ours. Since \cite{dun} have shown that galaxies at higher redshift have larger dust masses, the larger amount of dust in our sample is consistent with the trend observed for galaxies at $z <$ 0.5.

Another reason for the different dust mass distribution observed is due to the selection criterion. Selecting galaxies at 250 $\mu$m with robust detection at 3.6 $< \lambda <$ 500 \mic\ we can constrain both the warm and the cold component of the dust, while the IRAS selection is mostly sensitive to the warmest component. Some of the galaxies in our sample do not host dust that is warm enough to be detected at 60 $\mu$m even thought they are dust rich.

We also note, in the left panel of Fig. \ref{mdsfr}, that at fixed SFR the dust mass estimated in our sample is systematically higher than both \cite{smi} and \cite{dac}. The discrepancies between our sample and \cite{dac} is again a consequence of their selection criterion. At fixed SFR, IRAS selects galaxies with a lower dust content, because it is sensitive only to the warm component. The higher dust mass with respect to \cite{smi} could seem counterintuitive because at higher redshift we should trace galaxies with higher dust content. In this case this result is due to the depth of our data. Moreover the {\texttt timelinefitter} methods used for the photometry extraction allowed a robust flux density estimation down to 20 mJy \citep{pap}. In this way, we can investigate galaxies with lower dust luminosities: both in \cite{smi} and \cite{dac} the median dust luminosity is 6.4 $\times$ 10$^{10}$ L$_\odot$, almost three times higher than our median value of $\sim 2.3\pm 0.3$ $\times$ 10$^{10}$ L$_\odot$.

The right panel of Fig. \ref{mdsfr} shows $M_{dust}/M_\ast$ vs sSFR for the main sample. The results are consistent with H-ATLAS 250, although in our case we observe a long tail in the bottom left part of the panel representing a galaxy population with low SFR and dust content that were not visible in the H-ATLAS survey. However, the bulk of our galaxies strictly follow the relation observed both in \cite{smi} and in \cite{dac} with a small offset towards the bottom right part of the panel due to the different selection criteria.

\begin{figure*} 
\begin{center} 
\includegraphics[clip=,width=.49\textwidth]{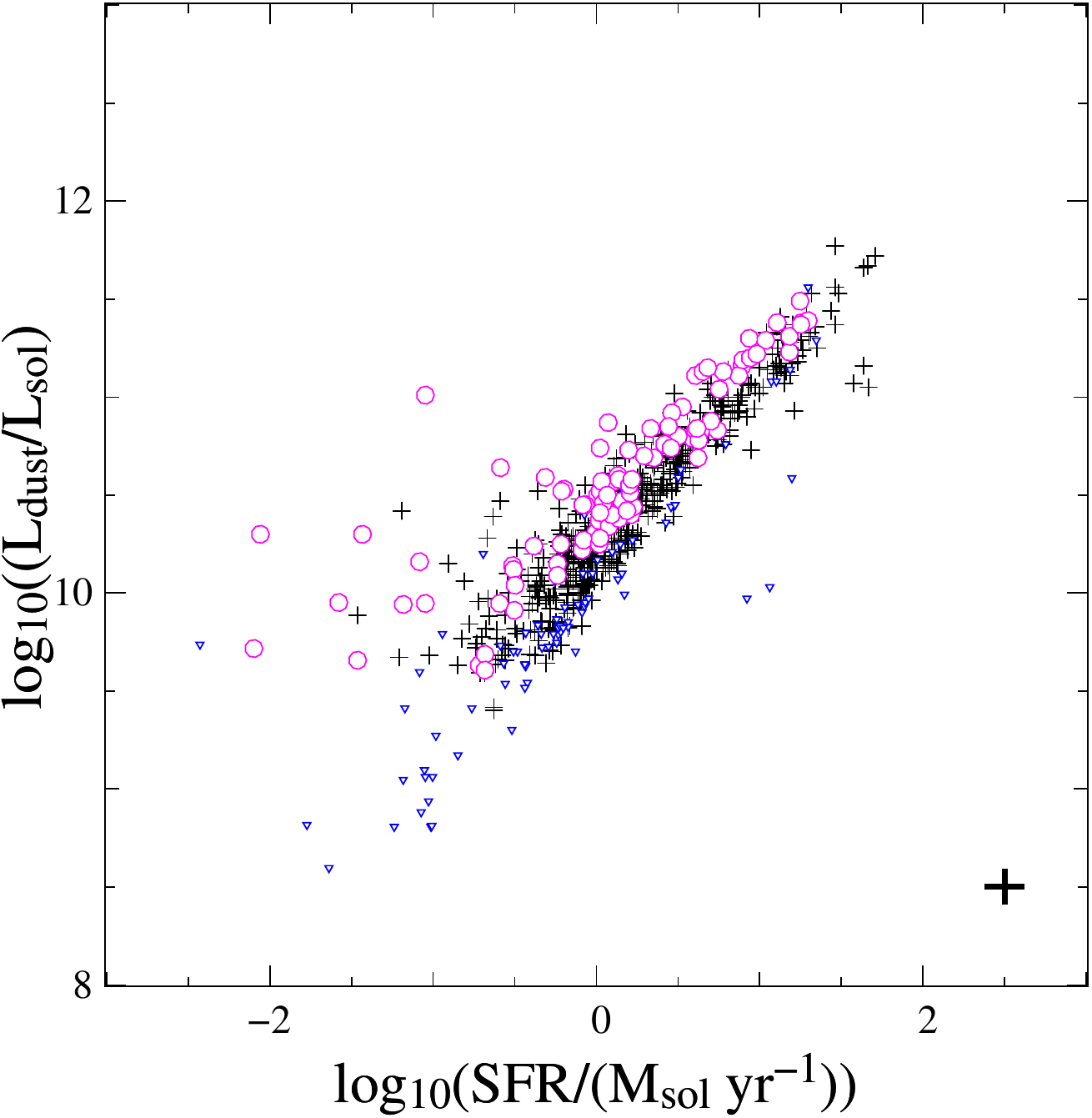}
\includegraphics[clip=,width=.49\textwidth]{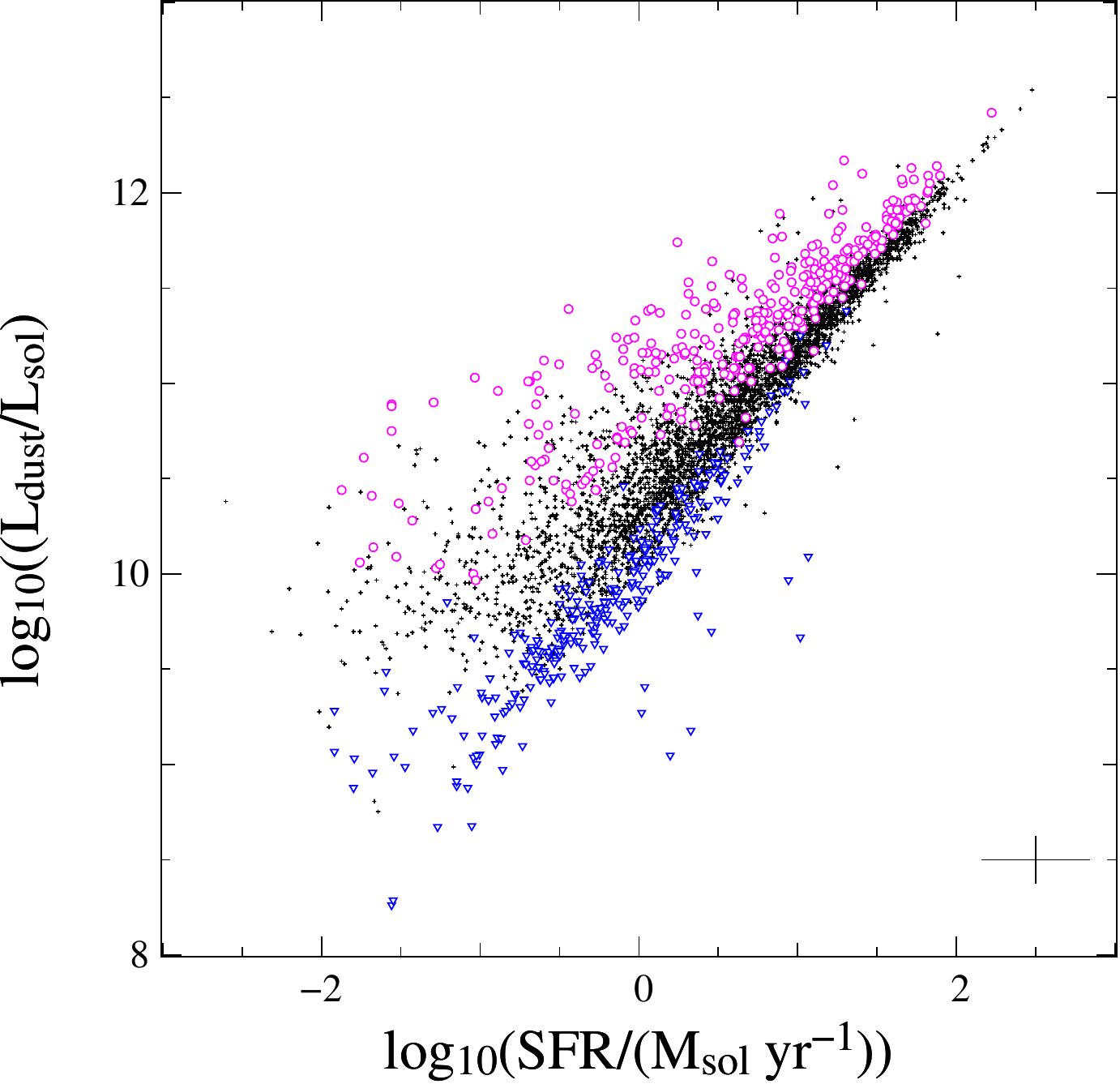}
\end{center} 
\caption{Dust luminosity ($L_{dust}$) vs SFR for the main (left) and extended sample (right) with average errors marked as a cross in the bottom right corner of each panel. Blue triangles and the magenta circles show two subsamples of approximatively the same size with $M_\ast$ $<$ 5 $\times$ 10$^9$ M$_\odot$ and $M_\ast$ $>$ 5 $\times$ 10$^{10}$ M$_\odot$ in the main sample, and $M_\ast$ $<$ 5 $\times$ 10$^9$ M$_\odot$ and $M_\ast$ $>$ 1.3 $\times$ 10$^{11}$ M$_\odot$ in the extended sample, respectively.}
\label{ldsfr} 
\end{figure*}

We now include, in the previous analysis, the stellar mass. In both panels of Fig. \ref{mdsfr} the blue triangles and the magenta circles show two subsamples of approximatively the same size with $M_\ast <$ 5 $\times$ 10$^9$ M$_\odot$ and $M_\ast >$ 5 $\times$ 10$^{10}$ M$_\odot$, respectively. We chose these values because in \cite{hea} $M_\ast \sim$ 1 $\times$ 10$^{10}$ M$_\odot$ corresponds to the threshold that discriminates two different populations: galaxies with masses lower than this value have a constant SFR in the last Gyr and give a strong contribution to the total SFRD, while for higher masses we found galaxies that are quenching their SFRs and are poor contributors to the SFRD. From previous considerations we expect that higher mass galaxies should show a different trend in the $M_{dust}$-SFR plane with respect to the low mass sample because the former occupy a different region in the $M_\ast$-SFR plane with respect to local main sequence galaxies (Figs. \ref{mseq} and \ref{mseq2}). However in Fig. \ref{mdsfr} this difference is not as clear: the two components are mixed and the only trend, as expected, is that galaxies with higher mass have on average higher dust content. Similar considerations are valid for the right panel of Fig. \ref{mdsfr}, where high mass galaxies tend to occupy the bottom left corner of the panel. This unclear evidence is due to the choice of the parameter to investigate the different trends of low and high mass galaxies. Dust mass includes both the warm and cold dust components and is not as closely linked to the SFR as the dust luminosity because in star forming regions dust is heated by young O-B stars that produce a more intense light. This means that dust luminosity is more sensitive to the emission of the warm dust component, and then more tightly related to the star formation process than the dust mass. 

We then studied the same relation shown in left panel of Fig. \ref{mdsfr} using the dust luminosity, $L_{dust}$, instead of $M_{dust}$ (see Fig. \ref{ldsfr}). The scatter in this relation, with respect to the relation that uses the dust mass (such as in Fig. \ref{mdsfr}), is strongly reduced, and the low mass subsample shows a stronger relation with the SFR than does the high mass component. Galaxies with mass below $M_\ast <$ 5 $\times$ 10$^9$ M$_\odot$ are still in an assembly phase up to $z$ = 0, populating in the SFR-$M_\ast$ relation the same region of the local main sequence spirals (magenta stars in Fig. \ref{mseq}). For this reason, in the left panel of Fig. \ref{ldsfr}, the low mass objects (blue triangles) have a tight correlation between $L_{dust}$ and SFR. Galaxies with higher mass, up to SFR $\sim$ 1 M$_\odot$ yr$^{-1}$, show a good correlation similarly to what is seen for low mass galaxies. However, at higher SFR high mass galaxies tend to deviate from this sequence and the correlation becomes more scattered, implying that for this population infrared luminosity is no longer a reliable tracer of the star formation process. To improve the statistical significance of the previous considerations we defined two homogeneous subsamples with $M_\ast <$ 5 $\times$ 10$^9$ M$_\odot$ and $M_\ast >$ 1.3 $\times$ 10$^{11}$ M$_\odot$ for the extended sample, recovering the same trend (right panel of Fig. \ref{ldsfr}).

Another interesting property of our sample is related to the dust-to-stellar mass ratio, which quantifies how dusty a galaxy is with respect to its stellar mass. \cite{dun} have clearly shown that galaxies at $z <$ 1 tend to be more dust rich at higher redshift. On the other hand, at $z <$ 1 most galaxies are already settled in the main sequence, and both the gas infall and the star formation are decreasing \citep{alm}. For this reason we expect, for galaxies at $z <$ 1, a lower growth rate than for their higher--z counterparts. During their assembly phase, galaxies show an approximately stable dust-to-stellar mass ratio, which reflects the higher star formation fuelled in turn by a higher gas infall rate. Once the peak of star formation is reached, the dust-to-stellar mass ratio decreases slowly. 

These simple considerations have been confirmed observationally. \cite{san} compared a sample of high redshift submillimeter galaxies (SMG) with the local galaxies of the Spitzer Infrared Nearby Galaxies Survey (SINGS) sample, finding in the former a decrease in the dust-to-stellar mass ratio of a factor 30. \cite{row} compared mass selected high redshift SMGs from \cite{mag} with a 250 $\mu$m--selected sample of galaxies at low redshift, taken from H-ATLAS. In this case, they found that SMGs have a dust-to-stellar mass ratio that is seven times higher than the low redshift sample. In both cases the high redshift galaxies have a dust-to-stellar mass ratio of 0.02. Surprisingly, in our low redshift sample the average dust-to-stellar mass ratio is $\sim$ 0.018, comparable with high redshift SMGs. 

\begin{figure} 
\begin{center} 
\includegraphics[clip=,width=.49\textwidth]{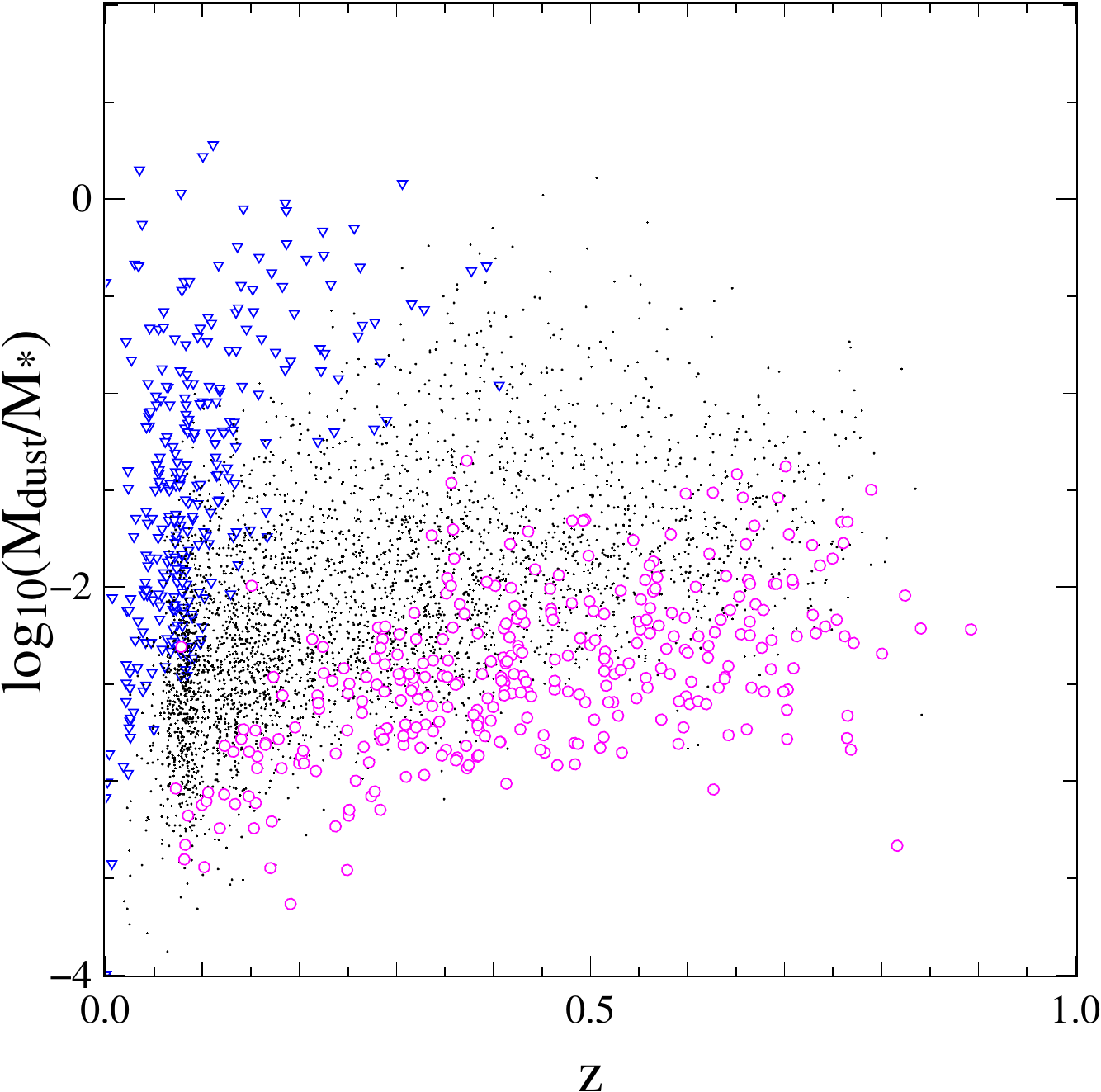}
\end{center} 
\caption{Dust-to-stellar mass ratio as a function of redshift for the extended sample (black dots). Blue triangles and magenta circles show the low and high mass subsample, as defined for Fig. \ref{ldsfr}.}
\label{dust_mstars} 
\end{figure}

To better understand this point, in Fig. \ref{dust_mstars} we compare the dust-to-stellar mass ratio of the extended sample as a function of redshift highlighting low mass (blue triangles) and high mass (magenta circles) galaxies. Low mass galaxies have an average dust--to--stellar mass ratio of 0.07, indicating a relatively high abundance of dust with respect to their mass, similar to high redshift SMGs. This confirms again that this subsample represents galaxies that are not quenching their star formation, they are instead evolving in a way similar to the local spirals and high redshift SMGs. The high mass subsample has an average dust-to-stellar mass ratio of 0.006, consistent with the results found by \cite{row} and \cite{san}, indicating for this population a decreased dust mass.

\begin{figure*}\begin{center} 
\includegraphics[clip=,width=.9\textwidth]{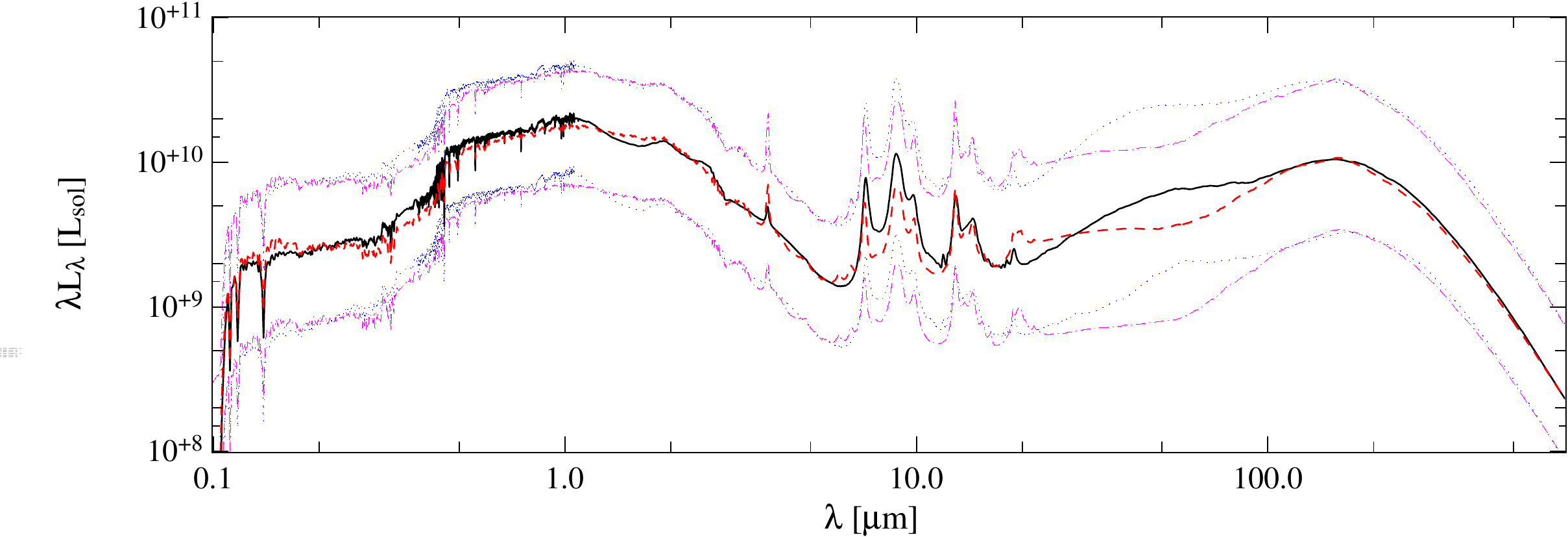}\end{center}
\caption{Total median spectra with the 16th and 84th percentile of the main sample obtained with MAGPHYS (black solid and blue dotted lines) and CIGALE (red dashed and magenta dot-dashed lines).}
\label{sedmagcig} 
\end{figure*}

\subsection{Towards a SED template}

The wavelength coverage of our sample is quite homogeneous and can be used to build a SED template representative of low FIR luminosity galaxies. We considered the SEDs obtained from the best fit of each galaxy of the main sample, and we built a median stacked spectra normalized to the average of $\lambda F_\lambda$ between 0.2 and 500 \mic. Fig. \ref{sedmagcig} compares the median spectra obtained with MAGPHYS and CIGALE. Overall there is a good agreement, confirming the goodness of our results and the consistency of the two methods. We see differences in regions where there are no anchor points to constrain the fit, between 25 $< \lambda <$ 100 $\mu$m. In this region MAGPHYS tends to predict higher IR emission with respect to CIGALE, but this difference does not affect the estimation of the dust luminosity, which is similar in the two methods, as shown in right panel of Fig. \ref{comp}. 

Hot dust emission at 25 $< \lambda <$ 100 $\mu$m is closely related to star formation, and it is due to a combination of warm dust and stochastically heated grains. The consistency of the dust luminosity distributions for both methods despite the mismatch seen at these wavelengths could be due to the different methods used to handle the stochastically heated dust grains. However, in this wavelength range there are no observational constraints, and so it is hard to distinguish between the uncertainties in the fit and/or possible variations in the physical properties of the dust grains in the sample. Moreover, the higher luminosity between 25 $< \lambda <$ 100 $\mu$m found in MAGPHYS could imply a further dust component at a different temperature, a possibility investigated in other works \citep[e.g.][]{cle}. Finally, as underlined in \cite{hay}, the phenomenological approach of \cite{dac2} reduce the efficacy of the dust emission model in MAGPHYS at observationally unsampled wavelengths. In their work \cite{hay} encourage using ``a more physically motivated model for dust emission, such as that of \cite{dra} to alleviate this problem \citep{cie}''. For these reasons we adopted a conservative approach, and decided to refer, for our analysis, to the SEDs obtained with CIGALE only as it uses a single dust component fitted by a \cite{dra} model, without any assumption about features between 22 $< \lambda <$ 100 $\mu$m.

\begin{figure*}\begin{center} 
\includegraphics[clip=,width=.9\textwidth]{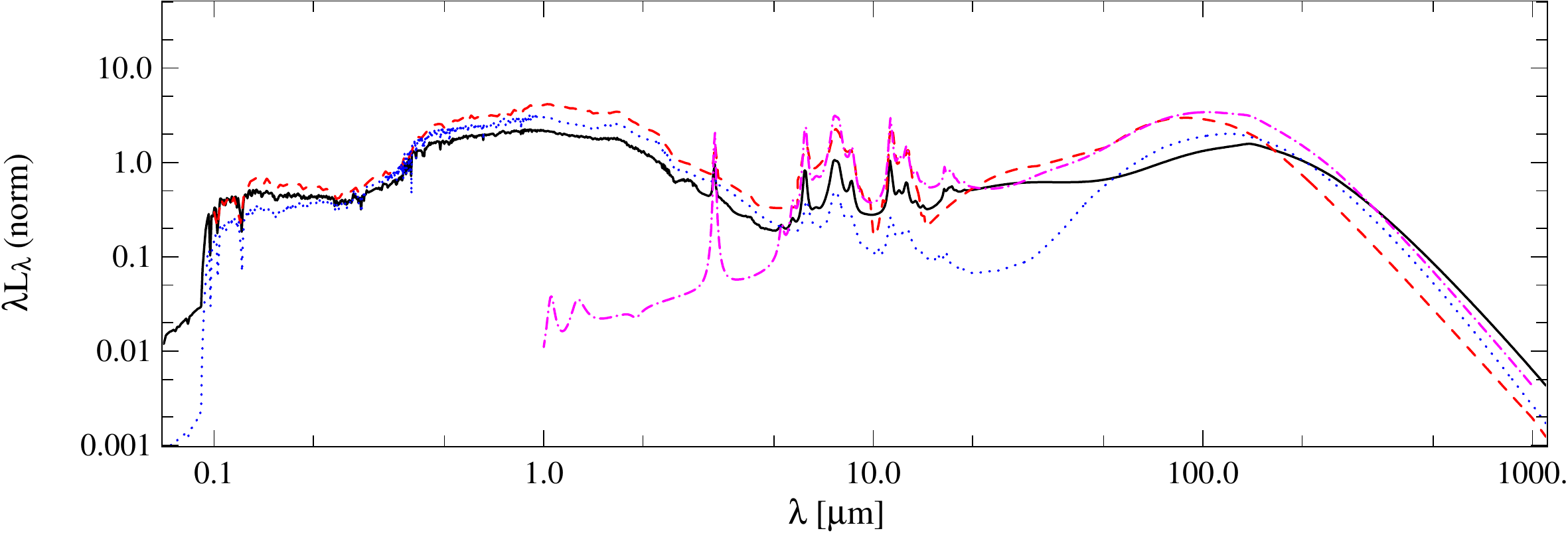}
\includegraphics[clip=,width=.9\textwidth]{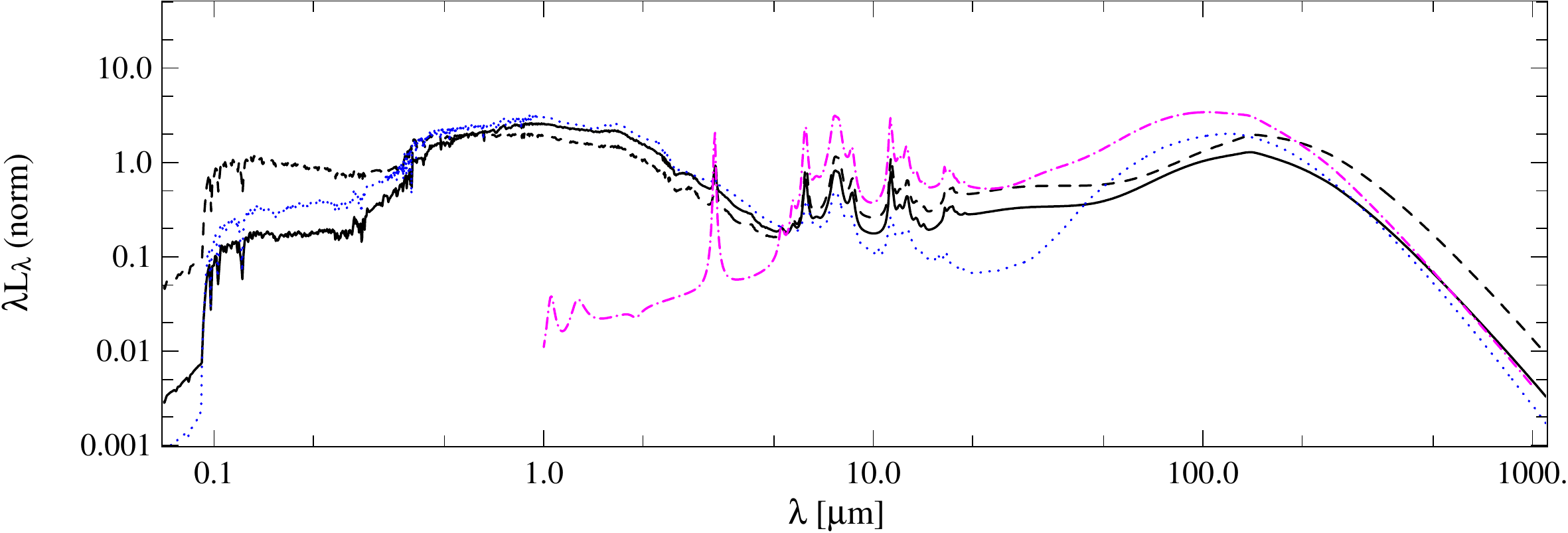}
\end{center}
\caption{Top panel: Normalized main sample median spectrum (black solid line) compared with the templates of \citealt{char} (red dashed line), \citealt{smi} (blue dotted line), and \citealt{cie} (magenta dot-dashed line). Bottom panel: Normalized main sample median spectrum obtained with CIGALE for high mass (black solid line) and low mass (black dashed line) subsamples, compared with the templates of \citealt{smi} (blue dotted line), and \citealt{cie} (magenta dot-dashed line).}
\label{template} 
\end{figure*}

In the top panel of Fig. \ref{template} our stacked SED is compared to different empirical templates based on low redshift galaxies. \cite{char} built a template based on $\sim$ 100 galaxies selected from a variety of published surveys. Objects were selected heterogeneously at different wavelengths, from 0.44 \mic\ to 850 \mic\ (Submillimeter Common-User Bolometer Array, SCUBA). In the wavelength range where the emission is dominated by the stellar component, despite the different offset due to the normalization, the shape of the spectra are similar. Above 10 \mic, where the infrared emission due to the dust is higher, the discrepancies increase, with dust thermal peak shifted at lower wavelengths, implying higher average temperatures.

A second template spectrum is obtained from H-ATLAS in \cite{smi}, improved by an order of magnitude with respect to the initial 1400 sources reported in their paper (Smith, private communication). Our median spectrum is almost identical to their template, if we ignore the region between 9 $ < \lambda <$ 100 \mic, despite the differences in star formations, stellar masses, and redshift range shown in Table \ref{tabella}. 

The shape of the median SEDs changes according to the selection criteria defined to built them. However, in \cite{smi}, it is shown that the parameter that mainly affects the shape of the median SEDs is the choice of dust luminosities. Their redshift distribution is higher, but despite this at fixed luminosity the median spectra in the available photometric bands are consistent. The main differences are in the spectral bands covered by WISE where \cite{smi} do not have any constraints. This evidence confirms the importance of MIR emission in building reliable SED templates for galaxies. The emission at MIR are on average lower than our sample when we have constraints in \hers\ bands \citep{smi}, and higher when there have no constraint in FIR \citep{char}.

A last template is taken from \cite{cie}, and considers about 150 local gas rich galaxies taken from the \hers\ Reference Survey \citep{bos2}. These galaxies are local star forming objects, and also in this case the peak of dust emission is at lower wavelengths, implying higher temperatures. With respect to the \cite{smi} template, this SED is more similar to the one derived for our sample at WISE bands, with a slightly reduced emission. These comparisons indicate that the low FIR luminosity galaxies have higher emission in the MIR wavelengths than previously thought, and models without constraints in these spectral regions \citep{smi} tend to recover lower emission. 

A point to note is that PAH emissions between 9-12 $\mu$m in local galaxies \citep{cie} are more pronounced than our sample because of their relatively higher SFR. However, even at our low SFR regimes, we observe emission in this range higher than the prediction of \cite{smi}.

The differences in this crucial region of galaxy SEDs are related to the galaxy stellar mass. In the bottom panel of Fig. \ref{template} we build the median SEDs of the main sample as estimated with CIGALE, taking into consideration the high mass (solid line) and low mass (dashed line) subsamples defined in Sect. \ref{pfir}. Low mass objects have higher emission in the UV and stronger PAH features with respect to the high mass subsample, implying higher SFRs. It is important to remember that we are dealing with normalized spectra, weighted for the light emitted mostly by dust and evolved stars (between 0.2 and 500 \mic). The bottom panel of Fig. \ref{template} indicates that low mass galaxies produce more stars than high mass galaxies with respect to their dust luminosity, and not an overall higher star formation, which would be false. We then find the same trend observed in Fig. \ref{ldsfr}, where at fixed dust mass we observed a scatter in the $L_{dust}$-SFR relation for high mass galaxies. They have higher dust content with higher temperatures, as witnessed by the peak of the dust emission shifted towards lower wavelengths, but their star formation is in a quenching phase, indicated by a lower UV emission.

\section{Conclusions}
\label{conc}

We have investigated the evolution of the SFRD at $z <$ 0.5 focusing on low FIR luminosity galaxies, which are the main contributors to the SFRD at these redshifts. To achieve this goal, we have considered the 250 $\mu$m--selected point source catalog presented in \cite{pap}, extracted from HeViCS \citep{dav2,dav}. We added complementary data to perform a multiwavelength analysis using different SED fitting techniques (MAGPHYS and CIGALE), and considered two different samples: a main sample with full spectral coverage and an extended sample in which we removed the constraints on the UV and the NIR data. We analyzed the correlation between SFR, $M_\ast$, and dust mass for a galaxy population with low dust content and low stellar masses, characterizing the bulk of the SFRD at low redshift. With the best fit SEDs obtained from the fit, we built a SED template representative of low FIR luminosity objects. The main results of our work are as follows:

\begin{itemize}

\item the main sample is formed of galaxies with moderate star formation and $M_\ast\sim 10^{10}$ M$_\odot$ occupying in a SFR-$M_\ast$ plane a region close to local spirals. Another component, with log$_{10}$(SFR/(M$_\odot$ yr$^{-1}$)) $>$ 0.5 and similar stellar masses has physical properties intermediate between the local spirals and higher redshift star forming galaxies. We introduced an extended sample with the UV and NIR constraints removed to tackle the population of dusty galaxies with low UV luminosities, that are also feeding the SFRD at $z <$ 0.5. Most of galaxies with $M_\ast > 3 \times 10^{10}$ M$_\odot$ are dusty galaxies in which the star formation is decreasing, while low mass galaxies ($M_\ast < 1 \times 10^{10}$ M$_\odot$) are mostly settled in the main sequence with $M_\ast$ and SFR consistent with local spirals.

\item Galaxies with $M_\ast <$ 5 $\times$ 10$^9$ M$_\odot$ populating, in the SFR-$M_\ast$ plane, the same region as local main-sequence spirals, show a tight correlation between $L_{dust}$ and SFR (Fig. \ref{ldsfr}). Galaxies with higher mass up to SFR $\sim$ 1 M$_\odot$ yr$^{-1}$ show a good correlation like that seen for low mass galaxies. However, at lower SFR, high mass galaxies tend to deviate from this sequence and the correlation becomes more scattered, implying that for this population infrared luminosity is no longer a reliable tracer of the star formation process.

\item investigating the dust-to-stellar mass ratio as a function of redshift we find that low mass galaxies have an average dust-to-stellar mass ratio similar to high redshift SMGs. This indicates that this subsample is representative of galaxies that are not quenching their star formation and are evolving in a way similar to the local spirals and high redshift SMGs.

\item we built a median stacked SED template representative of low FIR luminosity galaxies that we compared to different previous studies. Low FIR luminosity galaxies have higher emission in the MIR wavelengths than predicted by previous models without constraints at these wavelengths.

We investigated the differences in the SEDs of high and low mass subsamples. Normalizing to the dust emission, the low mass subsample has higher emission in the UV and higher PAH emission with respect to high mass galaxies, implying higher SFRs. High mass galaxies have higher dust content and lower UV emission as a consequence of the reduced star formation.
 
\end{itemize}

Low mass galaxies at low redshift are in an assembly phase and populate the main sequence diagram consistently with local spirals. This is in agreement with the recent claim of \cite{pen}, which in their ``gas regulator model'' emphasize that ``gas rich low-mass galaxies and dwarf galaxies are very unlikely to live around the equilibrium state at any epoch''. Low mass galaxies reach their equilibrium when settling into the main sequence: before this moment they show a tight correlation between dust luminosity and SFR as a consequence of the continuous recycling of processed and pristine gas from the cosmic web. From this point of view the main sequence does not seem to be the final stage of a virialization process of the halo in which the galaxy is forming, but an unstable stage of galaxy evolution that precedes the star formation quenching. The empirical relation for main sequence galaxies is a statistical effect due to the initial mass distribution of the dark matter halos where galaxies form. We are aware of the speculative nature of these considerations, because of the lack of information. For example, a proper characterization of the infall rate in low mass dark matter halos at higher redshift would be necessary, a possibility that will only be available in the future with the Square Kilometer Array (SKA). Another limit is due to the poor knowledge of the gas fraction in low mass galaxies. Previous works have mainly focused on IR detected galaxies in order to pledge a detection in a relatively small amount of time \citep{gea,bau}. However, we have shown that galaxies with low SFR and low FIR luminosities can occupy the same region of local spirals in the SFR-$M_\ast$ plane, indicating the presence of gas fueling the star formation process.

\begin{acknowledgements}
We warmly thank the anonymous referee for the suggestions. We thank also D. Munro for freely distributing his Yorick programming language (available at \texttt{http://www.maumae.net/yorick/doc/index.html}).
C. P. was also supported by PRIN-INAF 2009/11 grant (extended to 2012).
C.P. acknowledges support from the Science and Technology Foundation (FCT, Portugal) through the Postdoctoral Fellowship SFRH/BPD/90559/2012, PEst-OE/FIS/UI2751/2014, PTDC/FIS-AST/2194/2012, and through the support to the IA activity via the UID/FIS/04434/2013 fund.

Funding for SDSS-III has been provided by the Alfred P. Sloan Foundation, the Participating Institutions, the National Science Foundation, and the U.S. Department of Energy Office of Science. The SDSS-III web site is http://www.sdss3.org/.

SDSS-III is managed by the Astrophysical Research Consortium for the Participating Institutions of the SDSS-III Collaboration including the University of Arizona, the Brazilian Participation Group, Brookhaven National Laboratory, Carnegie Mellon University, University of Florida, the French Participation Group, the German Participation Group, Harvard University, the Instituto de Astrofisica de Canarias, the Michigan State/Notre Dame/JINA Participation Group, Johns Hopkins University, Lawrence Berkeley National Laboratory, Max Planck Institute for Astrophysics, Max Planck Institute for Extraterrestrial Physics, New Mexico State University, New York University, Ohio State University, Pennsylvania State University, University of Portsmouth, Princeton University, the Spanish Participation Group, University of Tokyo, University of Utah, Vanderbilt University, University of Virginia, University of Washington, and Yale University. 

This publication makes use of data products from the Two Micron All Sky Survey, which is a joint project of the University of Massachusetts and the Infrared Processing and Analysis Center/California Institute of Technology, funded by the National Aeronautics and Space Administration and the National Science Foundation.

We thank all the people involved in the construction and the launch of Herschel. SPIRE has been developed by a consortium of institutes led by Cardiff University (UK) and including Univ. Lethbridge (Canada); NAOC (China); CEA, LAM (France); IFSI, Univ. Padua (Italy); IAC (Spain); Stockholm Observatory (Sweden); Imperial College London, RAL, UCL-MSSL, UKATC, Univ. Sussex (UK); and Caltech, JPL, NHSC, Univ. Colorado (USA). This development has been supported by national funding agencies: CSA (Canada); NAOC (China); CEA, CNES, CNRS (France); ASI (Italy); MCINN (Spain); SNSB (Sweden); STFC and UKSA (UK); and NASA (USA). HIPE is a joint development (are joint developments) by the Herschel Science Ground Segment Consortium, consisting of ESA, the NASA Herschel Science Center, and the HIFI, PACS and SPIRE consortia.
\end{acknowledgements}


\end{document}